\documentclass[%
reprint,
superscriptaddress,
showpacs,
nofootinbib, nobibnotes, 
 amsmath,
 amssymb,
 aps,
 prl
]{revtex4-1}

\usepackage{times}

\usepackage{microtype}
\usepackage{blindtext}
\usepackage{dcolumn}
\usepackage{bm}
\usepackage{color, soul, colortbl} 
\usepackage[colorlinks=true,
						linkcolor=blue,
						urlcolor=blue,
						citecolor=blue,
						bookmarks=true,
						pdfborder={0 0 0}]{hyperref}

\usepackage{xcolor}
\usepackage{mdframed} 

\usepackage{graphicx,epsfig}
\usepackage{subfigure}
\usepackage{amsmath,amsfonts}
\usepackage{xfrac}
\usepackage{enumerate}
\usepackage[overload]{empheq}
\usepackage{mathtools}
\usepackage{cases} 
\usepackage{cancel} 

\begin{document}
\title{Interacting Discovery Processes on Complex Networks}

\author{Iacopo Iacopini}
\email[E-mail: ]{i.iacopini@qmul.ac.uk}
\affiliation{School of Mathematical Sciences, Queen Mary University of London, London E1 4NS, United Kingdom}
\affiliation{Centre for Advanced Spatial Analysis, University College London, London W1T 4TJ, United Kingdom}
\affiliation{The Alan Turing Institute, The British Library, London NW1 2DB, United Kingdom}
\author{Gabriele Di Bona}
\affiliation{School of Mathematical Sciences, Queen Mary University of London, London E1 4NS, United Kingdom}
\affiliation{Scuola Superiore di Catania, Universit\`a di Catania, Via Valdisavoia 9, 95123 Catania, Italy}
\author{Enrico Ubaldi}
\affiliation{Sony Computer Science Laboratories, 6 Rue Amyot, 75005 Paris, France}
\author{Vittorio Loreto}
\affiliation{Sony Computer Science Laboratories, 6 Rue Amyot, 75005 Paris, France}
\affiliation{Sapienza University of Rome, Physics Department, Piazzale Aldo Moro 5, 00185 Rome, Italy}
\affiliation{Complexity Science Hub (CSH) Vienna, A-1080 Vienna, Austria}
\author{Vito Latora}
\email[E-mail: ]{v.latora@qmul.ac.uk}
\affiliation{School of Mathematical Sciences, Queen Mary University of London, London E1 4NS, United Kingdom}
\affiliation{The Alan Turing Institute, The British Library, London NW1 2DB, United Kingdom}
\affiliation{Dipartimento di Fisica ed Astronomia, Universit\`a di Catania and INFN, I-95123 Catania, Italy}
\affiliation{Complexity Science Hub (CSH) Vienna, A-1080 Vienna, Austria}


\begin{abstract} 
	Innovation is the driving force of human progress. Recent urn models reproduce well the dynamics through which the discovery of a novelty may trigger further ones, in an expanding space of opportunities, but neglect the effects of social interactions. Here we focus on the mechanisms of collective exploration and we propose a model in which many urns, representing different explorers, are coupled through the links of a social network and exploit opportunities coming from their contacts. We study different network structures showing, both analytically and numerically, that the pace of discovery of an explorer depends on its centrality in the social network. Our model sheds light on the role that social structures play in discovery processes.
\end{abstract}
\maketitle

Discoveries are essential milestones for the progress of our societies~\cite{drews2000drug, erwin2004insights, wu2007novelty, sood2010interacting, perc2013self, rzhetsky2015choosing, fink2017serendipity, barron2018individuals, coccia2019nations, park2020novelty, hofstra2020diversity}. Recently, different mathematical approaches have been proposed to model the dynamics of innovation~\cite{cattuto2009collective, thurner2010schumpeterian, mcnerney2011role, dankulov2015dynamics, saracco2015innovation, andjelkovic2016topology, tadic2017mechanisms, Finkeaat6107, fink2019mathematical, coccia2019thetheory, pichler2020technological, ubaldi2020exploration}. Among these, of particular interest are those based on random processes with reinforcement~\cite{pemantle2007survey, launay2012generalized, aletti2020interacting}, such as P\`olya urns~\cite{hoppe1984polya,polya1930quelques}.
Urns have been extensively used to study and model a variety of systems and processes, from evolutionary economics, voting and contagions~\cite{simkin2011re, hayhoe2017polya1, hayhoe2018curing, berg1985paradox} to language and folksonomies \cite{gong2012studying, cattuto2007semiotic}.
More recently, they have been employed to filter information~\cite{marcaccioli2019polya} and grow social networks~\cite{ubaldi2020exploration}.
Interestingly, urns can also be used to model discovery processes, if opportunely combined with the concept of the \textit{adjacent possible} (AP)\textemdash {\it the set of all those things which are one step away from what is already known} (Kauffman~\cite{kauffman1996investigations}). 
This formulation of the AP, which dates back to concepts previously introduced by Farmer, Langton and others~\cite{packard1988adaptation, langton1990computation, langton2003artificial}, has been translated into the urn model with triggering (UMT), a particular process in which the space expands together with the discovery dynamics, and the appearance of a novelty opens up the possibilities of further discoveries~\cite{sood2010interacting, tria2014dynamics, loreto2016dynamics, gravino2016crossing, monechi2017waves}.
UMTs could successfully replicate the basic signatures of real-world discovery processes, such as the famous Heaps' and Zipf's laws~\cite{heaps1978information, zipf2016human}, often recurrent in complex systems~\cite{font2013scaling, perc2014matthew, dankulov2015dynamics, mazzolini2018heaps, mazzolini2018statistics, mazzolini2018zipf}, as well as Taylor's law~\cite{Tria_2018}.
It turns out that the Heaps' law, a sublinear growth of the number of distinct elements $D(t) \sim t^{\beta}$ with the number of elements $t$, well describes the pace at which scientists discover concepts, or users collect new items \cite{tria2014dynamics, iacopini2018network, mastrototaro2018mathematical}, with higher values of $\beta$ denoting a faster exploration of the AP.
However, despite the fact that existing models can capture essential underlying mechanisms behind the discovery of novelties, little emphasis is given to the collective dynamics of exploration and to the benefits that social interactions could bring. In fact, with the exception of Ref.~\cite{ubaldi2020exploration}, the modeled exploration dynamics refers to a single entity, representing, for example, the joint effort of researchers within a field~\cite{iacopini2018network}. Without taking into account the multiagent nature of the process, these models (\textit{i}) do not capture the heterogeneity of the pace of the individual explorers and (\textit{ii}) do not include the benefits brought by social interactions and collaborations. Indeed, empirical evidence of these mechanisms has been found in various contexts~\cite{salganik2006experimental, palovics2015temporal, ternovski2020social}, such as music listening, politics, voting, and language \cite{lazarsfeld1944people, bond201261, bryden2018humans}.\newline
\indent In this Letter, we propose a model of interacting discovery processes where an explorer is associated to each of the nodes of a social network
~\cite{albert2002statistical,newman2003structure,latora_nicosia_russo_2017}, and its dynamics is governed by an UMT.
Hence, the local dynamics of each node accounts for the presence of an AP, more precisely the {\it adjacent possible in the space of concepts.}
The social network makes the exploration a collective one, since processes of neighboring urns are coupled. This coupling expands the notion of the AP by adding a social dimension, represented by the set of opportunities one is possibly exposed to through his/her social contacts. We call this the {\it adjacent possible in the social space}.
Social networks have been extensively used as a substrate on top of which dynamical processes take place~\cite{porter2006dynamical, barrat2008dynamical}. Notice, however, that our setting crucially differs from the typical approach in which the network mediates, for example, the diffusion of innovations or social contagions~\cite{rogers2010diffusion, centola2018behavior}.
Here, the interactions among the many discovery processes reveals the twofold nature of the AP of each individual, highlighting the crucial role played by the social structure in determining the individual exploration dynamics.\newline 
%
%
%
%
\begin{figure}[t]
	\centering
	\includegraphics[width=\linewidth]{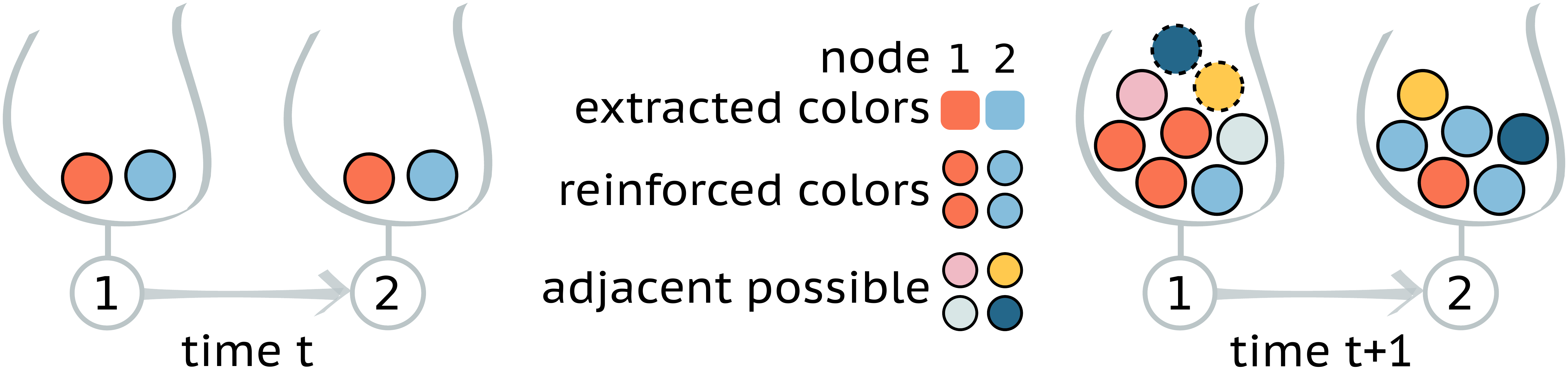}
	\vspace{-1.5em}
	\caption{\label{fig:Fig1}
		Illustration of the model in the case of a network with two nodes. Each node is equipped with an urn obeying the UMT with the same parameters $\rho=2$, $\nu=1$, and $M_0=\nu +1$. At the time $t$, the urns start with two balls, one red (R) and the other blue (B). Then, each node extracts a ball (1:R, 2:B), and therefore $\rho$ additional balls of the same colors are added to the respective urns (reinforcement). Also, since in both cases, the extracted balls represent a novelty for the respective nodes, $\nu+1$ balls of new colors are also added (adjacent possible). At $t+1$, node $1$ has access to all its balls plus two extra ones coming from the adjacent possible in the social space, i.e., the set of balls available through its neighbor (dashed borders).}
	\vspace{-1.5em}
\end{figure}
\indent \textit{Model.}\textemdash
Let us consider an unweighted directed graph $G(\mathcal{N},\mathcal{E})$, where $\mathcal{N}$ and $\mathcal{E}$ are, respectively, a set of $N=|\mathcal{N}|$ nodes and a set of $E=|\mathcal{E}|$ links. Each node of the graph represents an individual/agent, while link $(i,j)$ denotes the existence of a directed social relation from individual $i$ to $j$ (such that $i$ can benefit from $j$). The graph is  described by its adjacency matrix $\bm{A} \equiv \{a_{ij}\}$, whose element $a_{ij}$ is equal to 1 if link $(i,j)$ is present, and is 0 otherwise.
Each node $i$ is equipped with an UMT
that describes the discovery process of the agent $i$ \cite{tria2014dynamics}. We indicate the urn $i$ at time $t$ as $\mathcal{U}_{i}(t)$, while $\mathcal{S}_i(t)$ denotes the sequence of balls generated up to time $t$. Notice that $\mathcal{U}_{i}(t)$ is an unordered multiset of size $U_i(t)=|\mathcal{U}_{i}(t)|$, while $\mathcal{S}_i(t)$ is an ordered multiset of size $|\mathcal{S}_i(t)|=t$. 
Each urn $i$ is characterized by two 
parameters, $\rho_i$ and $\nu_i$. As in the original UMT, 
the {\it reinforcement}  parameter $\rho_i$ accounts for the number of balls of the same color that are added to the urn $i$ whenever a ball of a given color is extracted at time $t$.  
Furthermore, the {\it triggering} parameter $\nu_i$ controls the size of the {\it adjacent possible in the space of concepts}, as $(\nu_i + 1)$ balls of new colors are added to the urn of node $i$ whenever at time $t$ a color is extracted for the first time \cite{tria2014dynamics}.
In this abstract representation, the space of concepts\textemdash made by all the colors\textemdash expands in time together with each discovery process, without relying on a predefined structure~\cite{loreto2016dynamics}.
Discovery processes of different individuals are then coupled through the links of the network, representing social interactions. Namely, at each time $t$, the  individual $i$ draws a ball from an enriched urn, the so-called {\it social urn} of node $i$, $\tilde{\mathcal{U}}_i(t)$, composed by its own urn plus the additional balls present at time $t$ in the urns of its neighbors, without their reinforcement. The latter represents the AP in the social space. Figure \ref{fig:Fig1} illustrates the case of two nodes with a directed link.  
We thus have: 
\begin{equation}\label{eq:social_urn}
	\tilde{\mathcal{U}}_i (t)=\mathcal{U}_i (t) + \bigcup_{j \in\mathcal{N}} a_{ij}\mathcal{U}'_j(t) 
\end{equation}
where $\mathcal{U}'_j(t)=\mathcal{U}_j^{\left[ m=1 \right]} (t) \subseteq \mathcal{U}_j (t)$ is the underlying set of the multiset $\mathcal{U}_j (t)$ (with multiplicity $m=1$), i.e., the set of size $U'_j(t)=|\mathcal{U}'_j(t)|$ formed by its unique elements. Duplicates in the urn associated to node $j$ at time $t$ are indeed not considered. Thus, the ``memory" of node $j$ due to the reinforcement does not influence node $i$. Similarly, let us denote with $\mathcal{S}'_i(t)$ the underlying set of the sequence $\mathcal{S}_i(t)$, i.e., the sequence of all the unique elements of $\mathcal{S}_i(t)$. We consider synchronous updates for all the urns.\hfill \break 
%
%
\indent \textit{Pace of discovery.}\textemdash
As previous works have shown~\cite{tria2014dynamics}, the dynamics of novelties and innovations share a number of commonalities and can, thus, be thought as two sides of the same process; a novelty refers to the discovery of something by an individual (already known to others), while innovations are novelties that are new to everybody.
Here, we are interested in studying the asymptotic growth of the number of novelties\textemdash of each sequence\textemdash as a function of time (sequence length), representing the pace of discovery. 
We know, from standard results on the UMT \cite{tria2014dynamics}, that an isolated urn $i$ follows a Heaps' law, i.e., a power law behavior $D_{i}(t)\sim t ^{\beta_{i}}$~\cite{heaps1978information}, $D_{i}(t)=|\mathcal{S}'_i(t)|$ being the number of different elements contained in the sequence $\mathcal{S}_i(t)$ up to time $t$. Thus, the Heaps' exponent $\beta_i$ quantifies the speed at which the urn discovers new elements (by definition bounded by $\beta_i\leq1$).
Let us consider now a node $i$ that interacts through the network. In general, since $D_i(t)$ increases by one every time a ball is extracted for the first time, we can write $D_i(t+1)=D_i(t)+P^{\text{new}}_i(t)$, 
where $P^{\text{new}}_i(t)\in [0,1]$ is the probability that the ball extracted at node $i$ at time $t$ never appeared in $\mathcal{S}_i(t)$ before. In other words, $P^{\text{new}}_i(t)=\text{Prob}\left[D_i(t+1)=D_i(t)+1|D_i(t)\right]$ and we can express it as the fraction of discoverable balls over the total number of balls available to node $i$ at time $t$. This leads to an equation for the asymptotic Heaps' dynamics that in the continuous time limit reads:
\begin{equation}\label{eq:heaps_general}
	\frac{dD_i(t)}{dt} = P^{\text{new}}_i(t) = \frac{|\tilde{\mathcal{U}}_i(t)\ominus\mathcal{S}'_i(t)|}{\tilde{U}_i(t)} ,
\end{equation}
where $\mathcal{A} \ominus \mathcal{B}$ denotes the multiset obtained by removing all the elements in set $\mathcal{B}$ from the multiset $\mathcal{A}$ (all duplicates are removed).
Notice that if a node $i$ has an out-degree $\sum_j a_{ij}=0$, its associated Eq.~\eqref{eq:heaps_general} reduces to the one of an isolated urn, for which $\tilde{\mathcal{U}}_i (t)=\mathcal{U}_i (t)$. Thus, its Heaps dynamics for $\rho>\nu$ follows $D_i (t)\sim t^{{\nu}/{\rho}}$ for $t\to\infty$~\cite{tria2014dynamics, Tria_2018} (see Supplemental Material~\cite{SupMat}).
\nocite{tarjan1972depth, lu2016vital, Ide2014centrality,Ghosh2012centrality,katz1953}
%
In the most general case, where each node $i$ is equipped with a UMT($\rho_i, \nu_i$), the equation for the Heaps' laws of each node $i\in\mathcal{N}$ can be written as in Eq.~\eqref{eq:heaps_general}, by accounting for all the neighbors that are part of the social urn of node $i$. This can be done by using the non-zero elements of $\bm{A}$, so that the number of balls $\tilde{U}_i(t)$ in the social urn of node $i$ at time $t$ reads:
\begin{equation}\label{eq:tot_social_general}
	\tilde{U}_i(t) 
	=\rho_i t + \sum_{j\in\mathcal{N}} \big[a_{ij} + \delta_{ij}\big]\big[M_0 + (\nu_j+1)D_j(t)\big]
\end{equation}
where $M_0$ is the initial number of balls in each urn, and $\delta_{ij}$ stands for the Kronecker delta. Finally, the large time behavior of the number of different elements $D_i(t)$ for each node $i$ can be written as
{\small
	\begin{equation}\label{eq:heaps_explicit_general}
		\frac{dD_i(t)}{dt} =
		\frac{M_0\sum_{j}( a_{ij} + \delta_{ij}) + \sum_{j} \big[\delta_{ij}\nu_j + a_{ij}(\nu_j+1) \big]D_j(t)}{\rho_i t + \sum_{j} (a_{ij} + \delta_{ij})\big[M_0 + (\nu_j+1)D_j(t)\big]}.
\end{equation}}%
Equation~\eqref{eq:heaps_explicit_general} forms a system of $N$ coupled non-linear ordinary differential equations, with initial conditions $D_i(0)=0\ \forall i \in\mathcal{N}$, that can be numerically integrated for any network topology $\{a_{ij}\}$.\hfill \break
%
%
\begin{figure}
	\centering
	\includegraphics[width=0.9\linewidth]{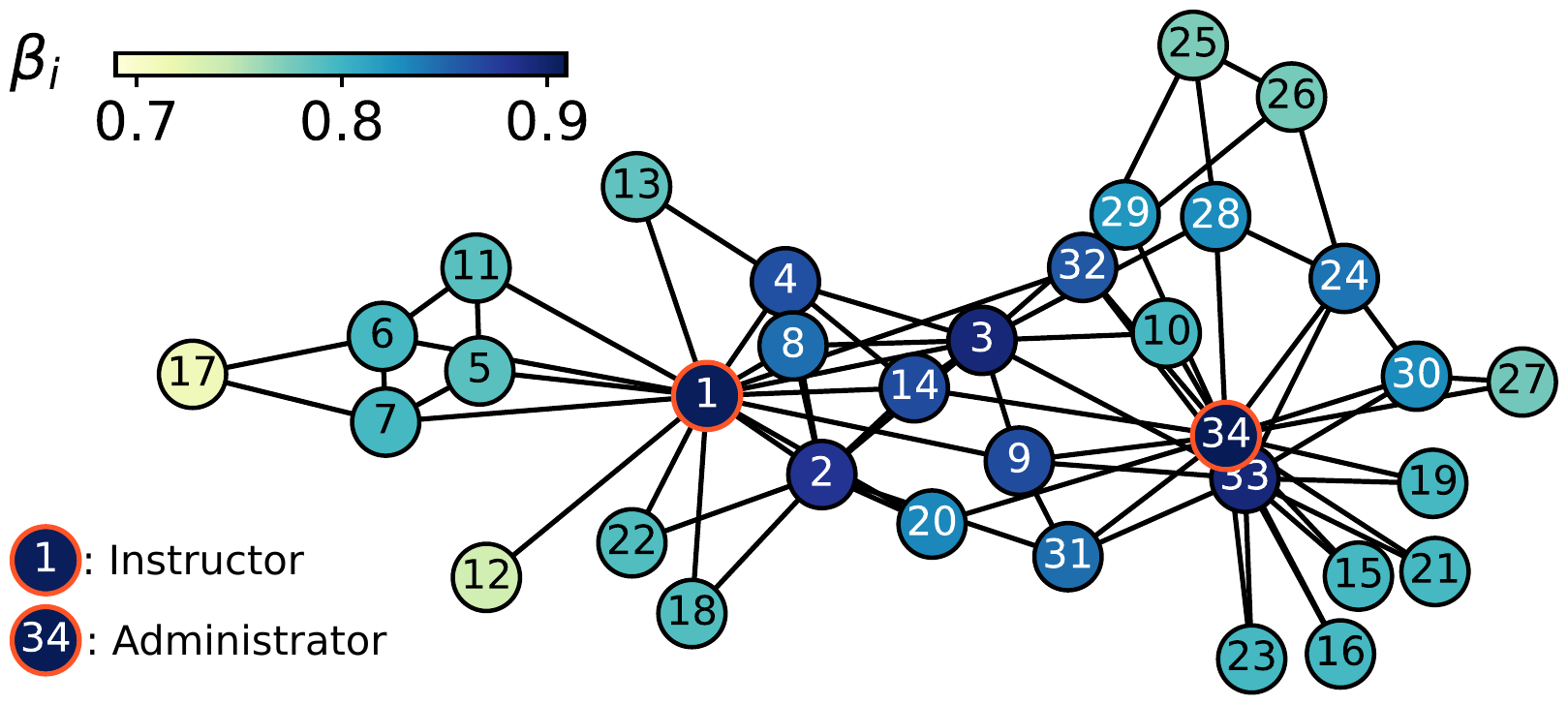}
	\vspace{-1em}
	\caption{\label{fig:Fig2} Dynamics of the interacting urns on the Zachary Karate Club network, whose nodes are colored according to the resulting Heaps' exponent.}
	\vspace{-1.5em}
\end{figure}
\indent \textit{Numerical results.}\textemdash
We start exploring the behavior of our model on the famous Zachary Karate Club network (ZKC) \cite{zachary1977information}, where each node is equipped with an UMT$(\rho=6,\, \nu=3)$ with same parameters and initial conditions. We run different simulations and observe, for each node $i$, the average growth of the number of distinct elements $D_i(t)$ as a function of time. We then extract the values of the Heaps' exponents of each node as $\beta_i=\beta_i(T)$, where 
$\beta_i(t)=\ln D_i(t)/\ln t$ and $T=10^4$. Figure \ref{fig:Fig2} shows the nodes of the networks colored accordingly. Notice the higher pace of discovery displayed by the notoriously central nodes corresponding to the instructor (node 1) and the administrator 
of the club (node 34). This proves that nodes with identical UMTs can have completely different dynamics, suggesting that a strategic location on the social network correlates with the discovery potential of an individual.
To further investigate this relation, we study the dynamics on five small directed networks. Figure~\ref{fig:Fig3}(a-e) shows the temporal evolution of $D_i(t)$ for each node $i$ of the networks displayed on the left. We report the simulated Heaps' laws (colored points), whose extracted exponents $\beta_i$ are shown in the legend. 
In addition, to assess the validity of Eq.~\eqref{eq:heaps_explicit_general}, we also plot the curves (continuous black lines) obtained using the appropriate $\{a_{ij}\}$. It can be seen that the analytical formalism introduced perfectly captures the Heaps' laws, since lines are almost indistinguishable from (simulated) points.
\begin{figure}[t]
	\centering
	\includegraphics[width=\linewidth]{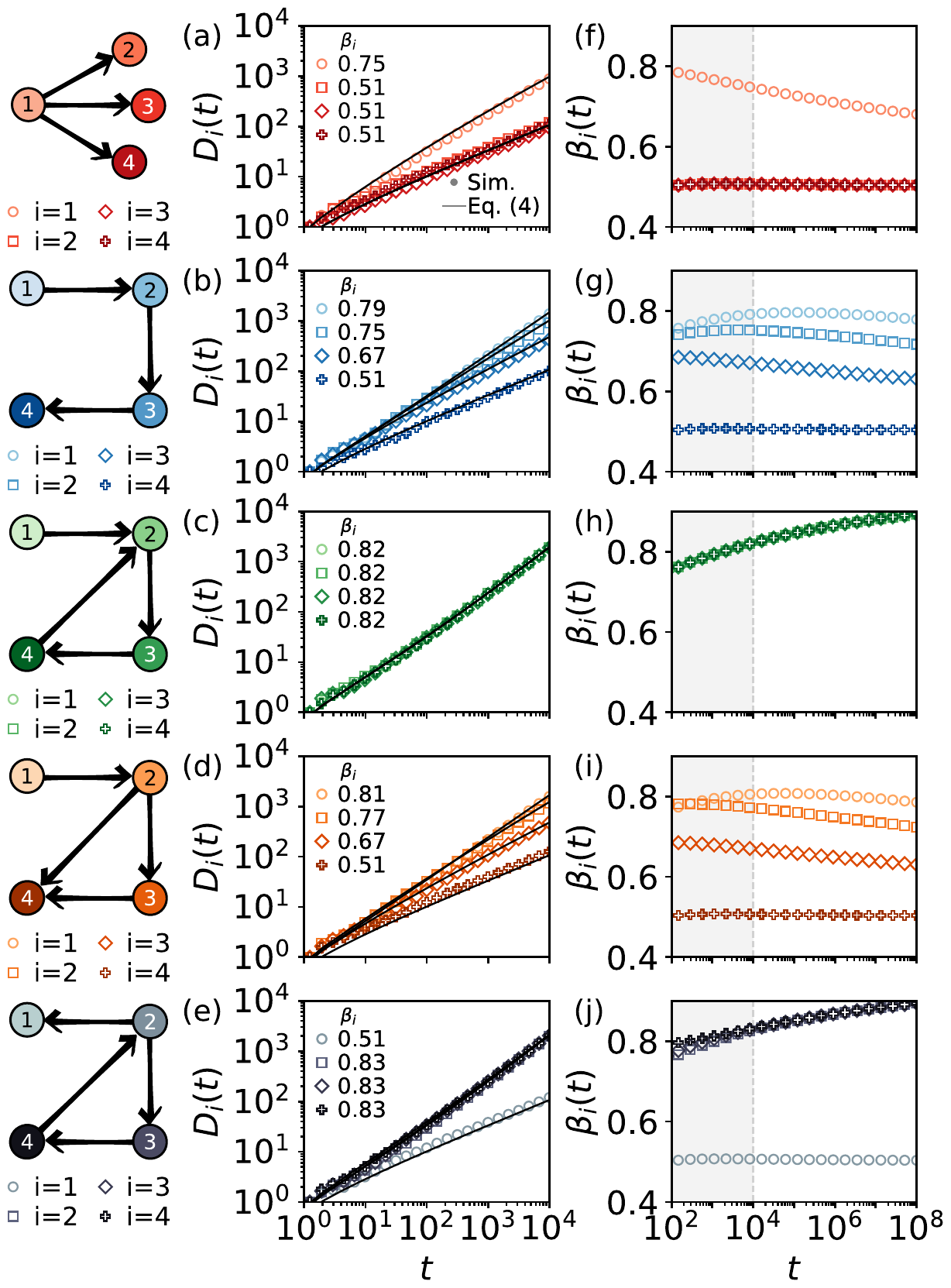}
	\vspace{-1.5em}
	\caption{\label{fig:Fig3} Heaps' dynamics of the interacting urns on five directed toy graphs (different symbols correspond to different nodes). Each node is equipped with an UMT with the same parameters $\rho=6$, and $\nu=3$. (a-e) Temporal evolution of the number of discoveries $D_i(t)$ for each node $i$ (associated Heaps' exponents $\beta_{i}$ in the legend). The solutions of Eq.~\eqref{eq:heaps_explicit_general}, shown as continuous black lines, are in perfect agreement with simulations. (f-j) Temporal behavior of the associated Heaps' exponents extracted at different times. The gray area up to $T=10^4$ corresponds to the values of (a-e).}
	\vspace{-1.5em}
\end{figure}
In particular, in Fig.~\ref{fig:Fig3}(a) we observe the highest pace of discovery in the node with more outgoing links. However, the non-trivial behaviors observed in Fig.~\ref{fig:Fig3}(b-e) for chains and graphs containing cycles indicate that the exponent of a node does not depend solely on local node properties. For instance, in Fig.~\ref{fig:Fig3}(d) node 2 has two outgoing links, while the others have one link only. In contrast with what is observed in Fig.~\ref{fig:Fig3}(a), here the highest pace of discovery is the one of node 1, whose social urn gets the benefits of the urn of node 2.
Moreover, in Figs.~\ref{fig:Fig3}(c) and (d) a simple change of direction of link $4\rightarrow2$ translates into completely different dynamics. We also notice that in both Fig.~\ref{fig:Fig3}(c) and (e) the presence of a cycle enhances the pace of discovery in a process of mutual exchange. However, while in Fig.~\ref{fig:Fig3}(d) node 1 is linked to the cycle and captures the same behavior of those in the cycle, in Fig.~\ref{fig:Fig3}(e) node 1 behaves as an individual urn.
We have further investigated whether the extracted $\beta_i$ may depend on the maximum time $T$ at which we have stopped the simulations. The curves reported in Fig.~\ref{fig:Fig3}(f-j) as a function of time for time up to $10^8$ clearly indicate that the systems, even for the small graphs considered, have not yet reached a stationary state. Thermalization times, that are typical of empirical trajectories of diffusion process~\cite{dosi2019dynamic}, here are strongly influenced by the topology of the network.
This can be seen by comparing the two $\beta_1(t)$ of Fig.~\ref{fig:Fig3}(f) and (g), both approaching\textemdash as we will see later\textemdash the asymptotic value $\nu/\rho=0.5$ but at very different timescales. Nevertheless, the ranking induced by the pace of discovery persists at all finite times.
In the next section we will further investigate this characteristic behavior, ultimately proving its universality for all networks (see Supplemental Material~\cite{SupMat}).

%
%
%
%
\indent \textit{Analytical results.}\textemdash 
In order to extract the asymptotic values of the Heaps' exponents, and their 
dependence on the network topology, we 
derive an analytical solution of Eq.~\eqref{eq:heaps_explicit_general} 
for $t\to \infty$. 
Let us suppose $\rho_i=\rho$ and $\nu_i=\nu\ \forall i \in \mathcal{N}$. For sufficiently high values of $\rho$ we have $\lim_{t\to\infty}D_i(t)/t=0\ \forall i$, so that the denominator of the rhs. of  Eq.~\eqref{eq:heaps_explicit_general} can be approximated by $\rho t$, leading to:
\begin{equation}\label{eq:approx_system_t}
	\frac{d\vec{D}(t)}{dt}\approx \frac{1}{t}\Bigg(\frac{\nu}{\rho}\bm{I}+\frac{\nu+1}{\rho}\bm{A}\Bigg)\vec{D}(t)
	= \frac{1}{t}
	\bm{M}\vec{D}(t),
\end{equation}
where $\vec{D}(t)\equiv \{D_i(t)\}_{i=1,\dots,N}$, $\bm{I}$ denotes the $N\times N$ identity matrix, and we have introduced the constant matrix $\bm{M}=f(\bm{A})=(\frac{\nu}{\rho}\bm{I}+\frac{\nu+1}{\rho}\bm{A})$. By operating the change of variable $t=e^z$, Eq.~\eqref{eq:approx_system_t} can be rewritten as $d_z\vec{D}(z)\approx \bm{M}\vec{D}(z)$, a standard first-order differential system, which leads to the solution
\begin{equation}\label{eq:approx_solution_t}
	\vec{D}(t)\approx \sum_{\ell=1}^{r}\sum_{p=0}^{m_\ell-1}\vec{c}_{p}\ln^{p}(t)\,t^{\lambda_\ell},
\end{equation}
where $\{\lambda_{\ell}\}_{\ell=1,\dots,r}$ and $\{m_\ell\}_{\ell=1,\dots,r}$ are the eigenvalues of $\bm{M}$ with their respective multiplicities, and $\vec{c}_{p}$ are vectors defined by the initial conditions. 
The asymptotic behavior of $D_i(t)$ is then governed by the leading term in  Eq.~\eqref{eq:approx_solution_t}, so that:  
\begin{equation}\label{eq:approx_leading_solution}
	D_i(t) \underset{t\to\infty}{\approx} u_i \ln^{\widehat{p}(i)}(t)\,t^{\widehat{\lambda}(i)}.
\end{equation}
where $\widehat{\lambda}(i)$ is the eigenvalue of $\bm{M}$ with the biggest real part such that the $i$-th entry of  at least one of its eigenvectors $\vec{c}_{p}$ is different from zero. 
Similarly, $\widehat{p}(i)$ is the maximum value of $p$ among these eigenvectors, 
and, in general, can be less than the multiplicity of the eigenvalue $\widehat{\lambda}(i)$ minus one.
For example, in the case of a chain as in Fig.~\ref{fig:Fig3}(b), the asymptotic solution is $D_i(t)\sim u_i \ln^{N-i}(t)\, t^{\nu/\rho}$.
In this example all the exponents tend to $\nu/\rho$ at large times, while at finite times nodes with higher powers in the logarithm show higher paces of discovery, thus explaining the behavior seen in Fig.~\ref{fig:Fig3}(g) (see Supplemental Material~\cite{SupMat}).
%
%
%

In the case of strongly connected graphs, Eq.~\eqref{eq:approx_leading_solution} simplifies: the logarithmic correction disappears and all the asymptotic exponents are equal to the maximum eigenvalue $\widehat{\lambda}= f(\widehat{\mu})$ of $\bm{M}$.
In fact, for the Perron-Frobeniusnius theorem~\cite{perron1907uber, frobenius1912matrizen}, $\bm{A}$ has a simple and positive maximum eigenvalue $\widehat{\mu}$ corresponding to an eigenvector $\vec{u}$ with all positive entries. Thus, 
the approximated solution becomes:
\begin{equation}\label{eq:approx_connected_solution}
	D_i(t) \underset{t\to\infty}{\approx} u_i \,t^{\widehat{\lambda}},
\end{equation}
where $u_i$ is proportional to the Bonacich eigenvector centrality~\cite{bonacich1972factoring} of node $i$, a global indicator of centrality that recursively quantifies the importance of a node from that of its neighbors, and not just from the number of neighbors.
\begin{figure}[t]
	\centering
	\includegraphics[width=\linewidth]{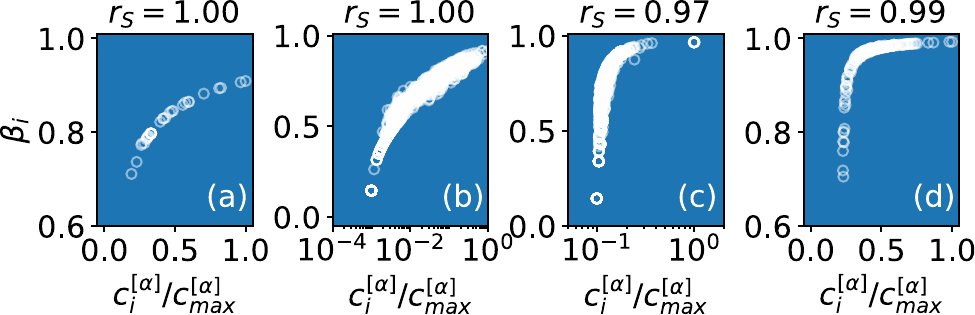}
	\vspace{-1.5em}
	\caption{\label{fig:Fig4} Scatter plot and Spearman's rank correlation coefficients $r_{S}$ between fitted Heaps' exponents $\beta_i$ and normalized $\alpha$-centrality $c^{[\alpha]}_i/c^{[\alpha]}_{\text{max}}$ associated to the $i=1,\dots,N$ nodes of four empirical networks.}
	\vspace{-1.5em}
\end{figure}
As a consequence of  Eq.~\eqref{eq:approx_connected_solution}, for strongly connected graphs every node has approximately the same behavior $t^{\widehat{\lambda}}$. What makes a node 
different from another is precisely the multiplicative factor $u_i$. 
In cycles and cliques, nodes are all structurally equivalent ($u_i =u ~ \forall i$), meaning that they all have the same $D_i(t)$. 
On the contrary, in graphs such as the ZKC (see Fig.~\ref{fig:Fig2}), the different values of $u_i$ play a very important role. Most central nodes, as the instructor and the chief administrator, are the fastest explorers (highest $\beta_i$), even having the same asymptotic Heaps' exponent $\widehat{\lambda}$.
%
%

In the more general case in which a graph is not strongly connected, Eq.~\eqref{eq:approx_leading_solution} still holds, and
the same argument can be applied to each of the strongly 
connected components to recursively 
find the values of $u_i$, $\widehat{p}(i)$, and $\widehat{\lambda}(i)$ (see Supplemental Material~\cite{SupMat}). In such cases, the eigenvector centrality needs to be replaced by its natural extension to non-strongly-connected graphs, i.e., the $\alpha$-centrality~\cite{bonacich2001eigenvector}.
We have investigated the correlation between 
the $\alpha$-centrality and the pace of discovery in real-world networks.
Figure \ref{fig:Fig4} shows the scatter plot of the number of discovered colors $D_i(T)$ 
and the normalized $\alpha$-centrality $c^{[\alpha]}_i/c^{[\alpha]}_{\text{max}}$ in four empirical social networks: (a) the ZKC \cite{zachary1977information}, (b) a Twitter network of followers~ \cite{de2010does}, (c) a co-authorship network in network science \cite{newman2006finding} and (d) a collaboration network between jazz musicians \cite{gleiser2003community} (see  Supplemental Material~\cite{SupMat}).
The high values of the Spearman's rank correlations ($r_{S}\geq 0.97$ in all cases) found in both undirected [Fig.~\ref{fig:Fig4}(a,c,d)] and directed networks [Fig.~\ref{fig:Fig4}(b)] is in agreement with our predictions. This confirms that, together with the AP in the space of concepts, it is crucial to take into account of  an AP in the social space. 

%
%

In conclusion, we have presented a first example in which stochastic (and not deterministic) processes are coupled over the nodes of a complex network, and 
analytical insights on the relations between structure and dynamics are possible.
The results highlight that the structural\textemdash not just local\textemdash properties of the nodes can strongly affect their ability to discover novelties. Our networked model of social urns is not just a simple extension of UMTs. What makes it novel and  different is the very same idea of coupling together many urns over a complex social network, and the concept of  ``social urn'' we have introduced. It is such a network coupling that spontaneously produces novel behaviors, such as different exponents of the Heaps' law in a single system, and has the potential to open new areas of research and applications.
This work represents only a first step toward the inclusion of structured interactions in discovery processes. Urns can, in fact, result oversimplified models for the dynamics of individual explorers.
Future works could consider non-identical urns, or even explore the effects of having individuals with a finite storage capacity, or where the adoption of the new might trigger the abandoning of the old, as for substitutive systems~\cite{jin2019emergence}.
Another natural extension would be considering discoveries and social relationships unfolding across different network layers~\cite{boccaletti2014structure} or higher-order structures~\cite{iacopini2019simplicial, battiston2020networks}.
In addition, it would be interesting to study the relationship with existing models of social spreading and meme popularity~\cite{gleeson2014competition, gleeson2016effects, d2019spreading}.
Finally, our results could be directly applied in studies on efficient team structures in cooperative creative tasks~\cite{schecter2018step, torrisi2019creative, monechi2019efficient, sinatra2016quantifying, fortunato2018science}.
%
%
\medskip
\begin{acknowledgements}
	I. I. and V. Lat. acknowledge support from EPSRC Grant No. EP/N013492/1. I. I. acknowledges support from the UK RDRF - Urban Dynamics Lab under the EPSRC Grant No. EP/M023583/1 and from The Alan Turing Institute under the EPSRC Grant No. EP/N510129/1. V. Lat. acknowledges support from the Leverhulme Trust Research Fellowship 278 ``CREATE: the network components of creativity and success". G. D .B. thanks Sony CSL, where part of this work was completed, for the kind hospitality. We thank U.~Alvarez-Rodriguez for the helpful comments and suggestions and V.D.P. Servedio and B. Monechi for interesting conversations about the early steps of this work.
	
	I. I. and G. D. B. contributed equally to this work.
\end{acknowledgements}


%


\clearpage
\setcounter{figure}{0}
\setcounter{table}{0}
\setcounter{equation}{0}
\makeatletter
\renewcommand{\thefigure}{S\arabic{figure}}
\renewcommand{\theequation}{S\arabic{equation}}
\renewcommand{\thetable}{S\arabic{table}}

\setcounter{secnumdepth}{2} 

\widetext
\begin{center}
	\textbf{\large Supplemental Material: Interacting Discovery Processes on Complex Networks}
\end{center}


\section{Analytical solutions}

In this section, we will study in more detail the analytical solutions we derived in the main text. We start by reviewing the case of an individual urn, which is equivalent to the urn model with triggering. We will then move on to more complicated cases, such as a pair of nodes, a chain, a cycle, a clique, ending with the formulation for the very general networks. Moreover, we will derive an algorithmic solution that allows deriving an analytical solution for each of the small networks studied in Fig.~3 of the main text. In every case, we will set the same parameters for each urn, so that $\rho_i=\rho$ (\textit{reinforcement}) and $\nu_i = \nu$ (\textit{triggering}) $\forall i = 1,\, \dots,\, N$. Each urn will be initialized with $M_0$ balls of different colors. These and the other colors\textemdash added from an individual $i$ when triggered by a discovery\textemdash will be taken from a single predefined set of discoverable balls of different colors. Notice that this set is shared by all the urns so that once a ball is drawn from an urn, it will not be available anymore to the others, except when enlarging the urn through the social adjacent possible (if they are connected).

\subsection{The single urn model}\label{sec:analytical_single_urn}

Let us consider the simplest case of an isolated urn, or equivalently, an urn on a node $i$ for which the out-degree $\sum_j a_{ij}$ is null. In this case, the dynamics will be the same of the Urn Model with Triggering (UMT)~\cite{tria2014dynamics,Tria_2018}, since the node does not have access to the balls of the neighbors, implying that its social urn will not be enriched ($\tilde{\mathcal{U}}_i (t)=\mathcal{U}_i (t)$). For such a node, the equation for the asymptotic Heaps' dynamics, Eq.~\eqref{eq:heaps_general} of the main text, reduces to:
\begin{equation}\label{eq:heaps_1node}
	\frac{dD(t)}{dt} = \frac{|\tilde{\mathcal{U}}(t)\ominus\mathcal{S}'(t)|}{\tilde{U}(t)} =\frac{U'(t)-D(t)}{U(t)} ,
\end{equation}
where $\mathcal{A} \ominus \mathcal{B}$ denotes the multiset obtained by removing all the elements in set $\mathcal{B}$ from the multiset $\mathcal{A}$ (all duplicates are removed).
Equation \eqref{eq:heaps_1node} can now be written as a function of the parameters of the model. In particular, we can write the total number of balls in the urn up to time $t$, $U(t)$, as the initial number of balls $M_0$, plus the $\rho$ balls added ($t$ times) as reinforcement, plus the $(\nu+1)$ balls added ($D(t)$ times, one for each novelty) due to the triggering mechanism:
\begin{equation}\label{eq:U_1node}
	U(t) = M_0+\rho t + (\nu+1)D(t).
\end{equation}
Similarly, the number of unique elements in the urn at time $t$, $U'(t)$, can be obtained by subtracting from $U(t)$ the $\rho t$ repeated balls coming from the reinforcement, that is: 
\begin{equation}\label{eq:U'_1node}
	U'(t)-D(t) = [U(t)-\rho t] -D(t) = M_0+\nu D(t).
\end{equation}
Thus, using Eq.~\eqref{eq:U_1node} and Eq.~\eqref{eq:U'_1node} in Eq.~\eqref{eq:heaps_1node} we obtain:
\begin{equation}\label{eq:heaps_1node_par}
	\frac{dD(t)}{dt} = \frac{M_0+\nu D(t)}{M_0+\rho t + (\nu+1)D(t)}.
\end{equation}
From now onwards we suppose that $t\gg M_0$, so that we can disregard $M_0$ in Eq.~\eqref{eq:heaps_1node_par} and in the similar equations we will obtain in the following sections. Therefore, after the introduction of the auxiliary variable $z(t)=\frac{D(t)}{t}$, Eq.~\eqref{eq:heaps_1node_par} can be rewritten as:
\begin{equation}
	\frac{dz(t)}{dt}t + z(t) = \frac{\nu z(t) t}{\rho t + (\nu+1)z(t) t},
\end{equation}
which can be integrated as:
\begin{equation}
	\int_{z(t_0)}^{z(t)} \frac{\rho + (\nu+1)z(t)}{z(t)[\nu-(\nu+1)z(t)-\rho]}dz(t) = \int_{t_0}^{t} \frac{1}{t}dt.
\end{equation}
The asymptotic solution ($t\to\infty$) depends on the parameters $\rho$ and $\nu$. It can be shown, as in the Supplemental Material of Ref.~\cite{tria2014dynamics, Tria_2018}, that the asymptotic solution for $D(t)$ is 
\begin{equation}\label{eq:sol_isolated_urn}
	\begin{cases}
		\rho>\nu & D(t) \sim (\rho - \nu)^\frac{\nu}{\rho}t^\frac{\nu}{\rho}\\
		\rho=\nu & D(t) \sim \frac{\nu}{\nu+1} \frac{t}{\ln t}\\
		\rho<\nu & D(t) \sim \frac{\nu - \rho}{\nu+1} t
	\end{cases}
\end{equation}
that is precisely the Heaps' law~\cite{heaps1978information}, with sublinear growth for $\rho>\nu$ and linear for the other cases. As empirical data has shown~\cite{heaps1978information, tria2014dynamics}, Heaps' laws usually have a sublinear behavior. For this reason, in this paper, we focus only on the case $\rho>\nu$.

\subsection{Two coupled urns}\label{sec:analytical_2_urns}

Let us consider now the simplest case of two coupled urns, that is a network with only two nodes connected by a directed edge $(1\rightarrow 2)$, as in Fig.~1 of the main text. This is equivalent to a directed chain of $N=2$ nodes, that will be discussed in the next Section for a general number $N$ of nodes.
The associated equations to determine the asymptotic Heaps' laws can be written expressing the probabilities $P^{\text{new}}_i(t)$ to draw a new ball as the the fraction of discoverable balls over the total number of balls available to node $i$ at time $t$:
\begin{subequations}\label{eq:heaps_2nodes}
	\begin{align}[left = \empheqlbrace\,]
		\frac{dD_1(t)}{dt} &= \frac{|\tilde{\mathcal{U}}_1(t)\ominus\mathcal{S}'_1(t)|}{\tilde{U}_1(t)}  \label{eq:heaps_2nodes_1}\\
		\frac{dD_2(t)}{dt} &= \frac{|\tilde{\mathcal{U}}_2(t)\ominus\mathcal{S}'_2(t)|}{\tilde{U}_2(t)} =\frac{U'_2(t)-D_2(t)}{U_2(t)}. \label{eq:heaps_2nodes_2}
	\end{align}
\end{subequations}
Notice that the right-hand side of Eq.~\eqref{eq:heaps_2nodes_2} is simplified since node $2$ does not have any outgoing link, and therefore its dynamics is the same of an isolated urn for which $\tilde{\mathcal{U}}_2 (t)=\mathcal{U}_2 (t)$. Thus, following the procedure discussed in the previous section, we have, for $\rho>\nu$:
\begin{equation}\label{eq:isolated_urn}
	D_2(t) \sim (\rho-\nu)^{\frac{\nu}{\rho}}t^{\frac{\nu}{\rho}}.
\end{equation}
The denominator $\tilde{U}_1(t)$ of Eq.~\eqref{eq:heaps_2nodes_1} can be expressed in terms of the two contributions coming from the two urns at time $t$, which reads:
\begin{equation}\label{eq:social_urn_2nodes_1}
	\begin{split}
		\tilde{U}_1(t) &=\overbrace{M_0 +\rho t + (\nu+1)D_1(t)}^{U_1(t)} + \overbrace{M_0 + (\nu+1)D_2(t)}^{U'_2(t)}\\
		&=2M_0 +\rho t + (\nu+1)\big[D_1(t)+D_2(t)\big].
	\end{split}
\end{equation}
Similarly, the numerator of Eq.~\eqref{eq:heaps_2nodes_1}, consisting in the number of balls present in the social urn of node $1$ at time $t$ which did not appeared yet in $\mathcal{S}_1(t)$, can be written as the total number of balls in the social urn of $1$ at time $t$, minus the number of duplicates, minus the number of balls that do not represent a novelty anymore with respect to the sequence $\mathcal{S}_1(t)$, i.e.:
\begin{equation}\label{eq:new_elements_2nodes_1}
	|\tilde{\mathcal{U}}_1(t)\ominus\mathcal{S}'_1(t)|=\tilde{U}_1(t)-\rho t - D_1(t).
\end{equation}
Then, using Eq.~\eqref{eq:social_urn_2nodes_1} and Eq.~\eqref{eq:new_elements_2nodes_1}, the final expression for Eq.~\eqref{eq:heaps_2nodes_1} reads:
\begin{equation}\label{eq:heaps_2nodes_explicit}
	\frac{dD_1(t)}{dt} = \frac{2M_0 + \nu D_1(t) + (\nu+1)D_2(t)}{2M_0 +\rho t + (\nu+1)\big[D_1(t)+D_2(t)\big]}.
\end{equation}
For large times ($t\gg M_0$) we can approximate Eq.~\eqref{eq:heaps_2nodes_explicit} as
\begin{equation}\label{eq:heaps_coupled_urn_explicit}
	\frac{dD_1(t)}{dt} \approx \frac{\nu D_1(t) + (\nu+1)D_2(t)}{\rho t + (\nu+1)\big[D_1(t)+D_2(t)\big]}.
\end{equation}
Let us assume now that the dynamics of node $2$ relaxes before the one of node $1$, so that we can solve Eq.~\eqref{eq:heaps_coupled_urn_explicit} independently from Eq.~\eqref{eq:isolated_urn}. In addition, if we suppose that $\lim_{t\to\infty}D(t)/t=0$, Eq.~\eqref{eq:heaps_coupled_urn_explicit} can be approximated as:
\begin{equation}\label{eq:heaps_2coupled_urns_approx}
	\frac{dD_1(t)}{dt} \approx \frac{\nu D_1(t)}{\rho t} + \frac{(\nu+1)D_2(t)}{\rho t}.
\end{equation}
The related homogeneous equation has a similar solution of Eq.~\eqref{eq:isolated_urn}, i.e.:
\begin{equation}\label{eq:hom_ass}
	\frac{d\overline{D}_1(t)}{dt} \approx \frac{\nu \overline{D}_1(t)}{\rho t}
	%
	%
	\implies
	\overline{D}_1(t) \sim (\rho-\nu)^{\frac{\nu}{\rho}}t^{\frac{\nu}{\rho}}.
\end{equation}
We now look for a solution like $D_1(t)=\kappa(t)\overline{D}_1(t)$ that, plugged into Eq.~\eqref{eq:heaps_2coupled_urns_approx}, leads to:
\begin{equation}
	\frac{d\kappa(t)}{dt}\overline{D}_1(t)+\kappa(t)\frac{d\overline{D}_1(t)}{dt}\approx\kappa(t)\frac{d\overline{D}_1(t)}{dt}+ \frac{(\nu+1)D_2(t)}{\rho t}.
\end{equation}
Thus, from Eq.~\eqref{eq:isolated_urn} and Eq.~\eqref{eq:hom_ass} we get:
\begin{equation}
	\frac{d\kappa(t)}{dt}=\frac{\nu+1}{\rho t}\frac{D_2(t)}{\overline{D}_1(t)}\approx\frac{\nu+1}{\rho t},
\end{equation}
whose solution is
\begin{equation}\label{eq:k_t}
	\kappa(t) \approx \frac{\nu+1}{\rho}\ln t.
\end{equation}
The asymptotic solution ($t\to \infty$) of $D_1(t)$ is then approximated by:
\begin{equation}\label{eq:solution_2_urns}
	D_1(t) \sim \frac{\nu+1}{\rho} (\rho-\nu)^{\frac{\nu}{\rho}}  \ln (t)\ t^{\frac{\nu}{\rho}}.
\end{equation}
In conclusion, comparing the solutions in Eq.~\eqref{eq:isolated_urn} and Eq.~\eqref{eq:solution_2_urns} the presence of an outgoing link increases the number of novelties with respect to an isolated urn dynamics. However, as we have shown here, this increase is approximately only logarithmic, meaning that we can see a slight increase at finite times which practically disappears for larger times. Le us also notice that this applies to the directed case, while in the case of an undirected link we would get identical Heaps' laws for both nodes $i=1,2$, without logarithmic corrections, but with higher exponents. This particular case is a cycle of two nodes, and as we will see in a dedicated section, cycles have their own behavior. 

\subsection{Chain of \texorpdfstring{$\bm{N}$}{\textit{N}} urns}\label{sec:analytical_N_chain}

Let us consider now a slightly more complicated case. Let us suppose that the network is composed by an open chain of $N$ urns, where there are only directed links $(i\rightarrow i+1)$
, with $i = 1,\,2, \dots,\, N-1$. This is the case considered in Fig.~3(b, g) of the main text, where in that case $N=4$.
Analogously to the previous case, the associated set of equations governing the growth of the number of novelties can be approximated to:
\begin{subequations}\label{eq:heaps_N_chain}
	\begin{align}[left = \empheqlbrace\,]
		\frac{dD_1(t)}{dt} &\approx \frac{\nu D_1(t) + (\nu+1)D_2(t)}{\rho t + (\nu+1)\big[D_1(t)+D_2(t)\big]}\label{eq:heaps_N_chain_1}\\
		\vdots \\  
		\frac{dD_{N-1}(t)}{dt} &\approx \frac{\nu D_{N-1}(t) + (\nu+1)D_N(t)}{\rho t + (\nu+1)\big[D_{N-1}(t)+D_N(t)\big]}\label{eq:heaps_N_chain_N-1}\\
		\frac{dD_N(t)}{dt} &\approx \frac{\nu D_N(t)}{\rho t + (\nu+1)D_N(t)}\label{eq:heaps_N_chain_N}
	\end{align}
\end{subequations}
We can solve the system by solving each equation, starting from the last one and recursively substituting its solution into the equation above. Indeed, since node $i=N$ does not have any outgoing link its independent Eq.~\eqref{eq:heaps_N_chain_N} can be immediately solved, resulting in the known asymptotic solution:
\begin{equation}\label{eq:sol_N_chain_N}
	D_N(t) \sim (\rho-\nu)^{\frac{\nu}{\rho}}t^{\frac{\nu}{\rho}}.
\end{equation}
As in the previous case, in Eq.~\eqref{eq:heaps_N_chain_N-1} we can consider $D_{N-1}(t)$ to be the only unknown variable. Then, following the same analytical steps presented in previous section leads to:
\begin{equation}
	D_{N-1}(t)\approx\frac{\nu+1}{\rho}(\rho-\nu)^{\frac{\nu}{\rho}}\ln{(t)}t^{\nu/\rho}.
\end{equation}
The same reasoning can be iterated for each node $i$. Let us now prove by induction on $i$ that the asymptotic solution is
\begin{equation}\label{eq:N_chain_induction_sol}
	D_i(t) = \frac{(\rho-\nu)^{\nu/\rho}}{(N-i)!}\left(\frac{\nu+1}{\rho}\ln(t)\right)^{N-i}t^{{\nu/\rho}}.
\end{equation}
We have already proved that this holds for $i= N$ and $i=N-1$. Let us now suppose that it holds for $i$ and let us prove it for $i-1$, with $1<i<N$. In the asymptotic limit, the equation for the growth of the number of novelties of node $i$ reads
\begin{equation}\label{eq:heaps_N_chain_i}
	\frac{dD_{i-1}(t)}{dt} \approx \frac{\nu D_{i-1}(t) + (\nu+1)D_i(t)}{\rho t + (\nu+1)\big[D_{i-1}(t)+D_i(t)\big]}.
\end{equation}
For the induction hypothesis, in Eq.~\eqref{eq:heaps_N_chain_i} the only unknown variable is $D_i(t)$. Therefore, we can consider the homogeneous associated equation
\begin{equation}
	\frac{d\overline{D}_{i-1}(t)}{dt} \approx \frac{\nu \overline{D}_{i-1}(t)}{\rho t},
\end{equation}
which provides the approximated solution:
\begin{equation}\label{eq:hom_ass_N_chain_i}
	\overline{D}_{i-1}(t) \approx (\rho-\nu)^{\frac{\nu}{\rho}}t^{\frac{\nu}{\rho}}.
\end{equation}
As for the case of two coupled urns, we now look for a solution like $D_{i-1}(t)=\kappa(t)\overline{D}_{i-1}(t)$, that, plugged into Eq.~\eqref{eq:heaps_N_chain_i}, leads to:
\begin{equation}
	\frac{d\kappa(t)}{dt}\overline{D}_{i-1}(t) + \cancel{\kappa(t)\frac{d\overline{D}_{i-1}(t)}{dt}} \approx \cancel{\kappa(t)\frac{d\overline{D}_{i-1}(t)}{dt}} + \frac{(\nu+1)D_i(t)}{\rho t}.
\end{equation}
Thus, we get
\begin{equation}
	\frac{d\kappa(t)}{dt} \approx \frac{\nu+1}{\rho t} \frac{D_{i}(t)}{\overline{D}_{i-1}(t)} \approx \frac{1}{(N-i)!}\frac{\nu+1}{\rho t} \left(\frac{\nu+1}{\rho}\ln(t)\right)^{N-i},
\end{equation}
whose solution is
\begin{equation}\label{eq:k_t_N_chain_i}
	\kappa(t) \approx \frac{1}{(N-(i-1))!}\left(\frac{\nu+1}{\rho}\ln(t)\right)^{N-(i-1)}.
\end{equation}
Finally, after combining Eq.~\eqref{eq:hom_ass_N_chain_i} and Eq.~\eqref{eq:k_t_N_chain_i}, we reach the solution for the dynamics of node $i-1$, that reads:
\begin{equation}
	D_{i-1}(t)\approx\frac{(\rho-\nu)^{\nu/\rho}}{(N-(i-1))!}\left(\frac{\nu+1}{\rho}\ln(t)\right)^{N-(i-1)}t^{\nu/\rho},
\end{equation}
which completes the proof by induction.

Finally, it is worth observing that the Heaps' laws would be very different if the links were undirected. This would indeed result, similarly to undirected cycles, in higher asymptotic Heaps' exponents.

\subsection{Cycle of \texorpdfstring{$\bm{N}$}{\textit{N}} urns}\label{sec:analytical_cycle}

\textit{\textbf {Directed cycle}}\textemdash
Let us consider the case of directed cycles. As we will see, this is the simplest system leading to asymptotic Heaps' exponents that are higher than that of an individual urn. Let us hence suppose that every node $i$ is connected just to the following one, node $i+1$, with a directed link $(i\rightarrow i+1)$,
with $i = 1,\,2, \dots,\, N$, where we identify node $N+1$ with node $1$.
For a generic node $i$, the asymptotic differential equation for the growth of the number of novelties reads:
\begin{equation}\label{eq:heaps_cycle_i}
	\frac{dD_i(t)}{dt} \approx \frac{\nu D_i(t) + (\nu+1)D_{i+1}(t)}{\rho t + (\nu+1)\big[D_i(t)+D_{i+1}(t)\big]}.
\end{equation}
For symmetry reasons, the dynamics of each node is the same, implying that $D_1(t) \approx \dots \approx D_i(t) \approx \dots \approx D_N(t)$. Hence, Eq.~\eqref{eq:heaps_cycle_i} becomes
\begin{equation}\label{eq:heaps_cycle_symmetric_i}
	\frac{dD_i(t)}{dt} \approx \frac{(2\nu+1)D_i(t)}{\rho t + 2(\nu+1)D_i(t)},
\end{equation}
that is equal to the equation of an individual urn [see Eq.~\eqref{eq:heaps_1node_par}], with $\nu' = 2\nu+1$. Therefore,
if $\rho>\nu'$ we have the solution
\begin{equation}\label{eq:sol_cycle_directed}
	D_i(t) \approx (\rho-2\nu-1)^{\frac{2\nu+1}{\rho}}t^{\frac{2\nu+1}{\rho}}. 
\end{equation}

\textit{\textbf {Undirected cycle}}\textemdash
Let us now consider cycles composed by undirected links. Let us suppose that $N>2$, considered that for $N = 1$ the network reduces to an individual urn, and for $N=2$ it is equivalent to a directed cycle of 2 nodes. For $N>2$,
each node $i$ is therefore connected to two different nodes $i-1$ and $i+1$, and the associated equations to be solved are:
\begin{equation}\label{eq:heaps_cycle_undirected_i}
	\frac{dD_i(t)}{dt} \approx \frac{\nu D_i(t) + (\nu+1)D_{i-1}(t)+ (\nu+1)D_{i+1}(t)}{\rho t + (\nu+1)\big[D_i(t)+(\nu+1)D_{i-1}(t)+D_{i+1}(t)\big]}.
\end{equation}
Again, for symmetry reasons, we can equivalently write Eq.~\eqref{eq:heaps_cycle_undirected_i} as
\begin{equation}\label{eq:heaps_cycle_undirected_symmetric_i}
	\frac{dD_i(t)}{dt} \approx \frac{(3\nu+2)D_i(t)}{\rho t + 3(\nu+1)D_i(t)},
\end{equation}
that is equal to the equation of an individual urn [see Eq.~\eqref{eq:heaps_1node_par}], with $\nu'' = 3\nu+2$. Therefore, if $\rho>\nu''$ we have the solution
\begin{equation}\label{eq:sol_cycle_undirected}
	D_i(t) \approx (\rho-3\nu-2)^{\frac{3\nu+2}{\rho}}t^{\frac{3\nu+2}{\rho}}.
\end{equation}
Notice that for undirected cycles, since all connections are mutual, the resulting paces of discovery are higher than those in the directed case. However, in both cases, directed and undirected, the dynamics of each node does not depend on the length of the cycle. 

\subsection{Clique of \texorpdfstring{$\bm{N}$}{\textit{N}} urns}\label{sec:approx_Heaps_clique}

Let us consider a $N$-clique, that is a fully connected network of $N$ nodes, equivalently directed or undirected. Being every node $i$ connected to all other nodes, all nodes are equivalent, and the general equation for the growth of the number of novelties of node $i$ reads: 
\begin{equation}\label{eq:heaps_clique_undirected_i}
	\frac{dD_i(t)}{dt} \approx \frac{\nu D_i(t) + (\nu+1)\sum_{j\neq i}D_j(t)}{\rho t + (\nu+1)\sum_{j=1}^N D_j(t)}.
\end{equation}
For symmetry reasons, each urn follows the same dynamics and we can equivalently write Eq.~\eqref{eq:heaps_clique_undirected_i} as
\begin{equation}\label{eq:heaps_clique_undirected_symmetric_i}
	\frac{dD_i(t)}{dt} \approx \frac{[N(\nu+1)-1]D_i(t)}{\rho t + N(\nu+1)D_i(t)},
\end{equation}
that is equal to the equation for an individual urn [see Eq.~\eqref{eq:heaps_1node_par}], with $\nu''' = N(\nu+1)-1$. Therefore, if $\rho>\nu'''$ we have the solution
\begin{equation}
	D_i(t) \approx (\rho-N(\nu+1)-1)^{\frac{N(\nu+1)-1}{\rho}}t^{\frac{N(\nu+1)-1}{\rho}}.
\end{equation}
Let us observe that for any network with $N$ nodes, the maximum allowed Heaps' exponent is hence $[N(\nu+1)-1]/\rho$, which occurs only in the case of a fully connected network.

\subsection{The general case}

Let us consider a general graph $G(\mathcal{N},\mathcal{E})$, either directed or undirected. 
In order to write and solve the equations for the growth of the number of novelties, we first have to calculate the probability $P_i^{\text{new}}(t)$ of drawing a new ball from the urn of each node $i$. This can be done by considering the number of different colors present in the social urn $\tilde{\mathcal{U}}_i(t)$ of node $i$ at time $t$ that have not been discovered yet by $i$, divided by the total number of balls $\tilde{U}_i(t)$ present in its social urn at that time.
The numerator can be expressed as $|\tilde{\mathcal{U}}_i(t)\ominus\mathcal{S}'_i(t)|$, which is the length of the multiset obtained by removing from the multiset $\tilde{\mathcal{U}}_i(t)$ all the elements appeared in the sequence (taking out all duplicates).
In other words, it is the number of unique colors present in the urn of node $i$ and in the one of its neighbors (without their multiplicity) minus the number of colors already drawn (unique elements in the sequence of $i$).
Considering that all (and only) the already discovered balls are those that have been reinforced and that the number of triggered colors added to the urn $j$ is exactly $(\nu+1)D_j(t)$, we can write:
\begin{equation}\label{eq:heaps_general_SM}
	\frac{dD_i(t)}{dt} = P^{\text{new}}_i(t) = \frac{|\tilde{\mathcal{U}}_i(t)\ominus\mathcal{S}'_i(t)|}{\tilde{U}_i(t)}
	= \frac{M_0 + \nu D_i(t) + \sum_{j \neq i} a_{ij}\big[M_0 + (\nu+1)D_j(t)\big]} {\rho t + M_0 + (\nu+1) D_i(t) + \sum_{j \neq i} a_{ij}\big[M_0 + (\nu+1)D_j(t)\big]} ,
\end{equation}
or, equivalently:
\begin{equation}\label{eq:heaps_general_SM2}
	\frac{dD_i(t)}{dt} = 
	\frac{M_0\sum_{j}( a_{ij} + \delta_{ij}) + \sum_{j} \big[\delta_{ij}\nu + a_{ij}(\nu+1) \big]D_j(t)}{\rho t + \sum_{j} (a_{ij} + \delta_{ij})\big[M_0 + (\nu+1)D_j(t)\big]}.
\end{equation}
For $t \gg M_0$ we can disregard the presence of $M_0$ in Eq.~\eqref{eq:heaps_general_SM2}. As shown above for $N$-cliques, in the asymptotic limit $t\to\infty$ the growth of the number of novelties obeys an Heaps' law with maximum exponent  $[N(\nu+1)-1]/\rho$. This means that if $\rho$ is high enough, we can approximate the denominator on the r.h.s. of Eq.~\eqref{eq:heaps_general_SM2} to $\rho t$. After finding the approximated solution, we will estimate the set of parameters for which this approximation is valid for any topology. Therefore, in the asymptotic limit and with a proper choice of the parameters, Eq.~\eqref{eq:heaps_general_SM2} can be rewritten as: 
\begin{equation}\label{eq:approx_heaps_general_SM}
	\frac{dD_i(t)}{dt} \approx \frac{\sum_{j} \big[\delta_{ij}\nu + a_{ij}(\nu+1) \big]D_j(t)} {\rho t},
\end{equation}
which can be expressed in a more compact way as:
\begin{equation}\label{eq:approx_heaps_general_matrix_SM}
	\frac{d\vec{D}(t)}{dt}\approx \frac{1}{t}\Bigg(\frac{\nu}{\rho}\bm{I}+\frac{\nu+1}{\rho}\bm{A}\Bigg)\vec{D}(t)
	= \frac{1}{t} \frac{f(\bm{A})\vec{D}(t)}{t}
	= \frac{1}{t} \bm{M}\vec{D}(t),
\end{equation}
where $\bm{I}$ is the $N\times N$ identity matrix and $\bm{M} = f(A)$, with $f(x) = \frac{\nu}{\rho}+\frac{\nu+1}{\rho}x$.
By operating the change of variable $t=e^z$, Eq.~\eqref{eq:approx_heaps_general_matrix_SM} can be rewritten as a standard first-order differential system, i.e. $d_z\vec{D}(z)\approx \bm{M}\vec{D}(z)$, which leads to the solution
\begin{equation}\label{eq:approx_solution_t_SM}
	\vec{D}(t) \approx \sum_{\ell=1}^{r}\sum_{p=0}^{m_\ell-1}\vec{c}_{p}\ln^{p}(t)\,t^{\lambda_\ell},
\end{equation}
where $\{\lambda_{\ell}\}_{\ell=1,\dots,r}$ and $\{m_\ell\}_{\ell=1,\dots,r}$ are the eigenvalues of $\bm{M}$ with their respective multiplicities, and $\vec{c}_{p}$ are vectors defined by the initial conditions. 
The asymptotic behavior of the number of novelties $D_i(t)$ discovered by node $i$ at time $t$ is then governed by the leading term in  Eq.~\eqref{eq:approx_solution_t_SM}, so that we can write:  
\begin{equation}\label{eq:approx_leading_solution_SM}
	D_i(t) \underset{t\to\infty}{\approx} u_i \ln^{\widehat{p}(i)}(t)\,t^{\widehat{\lambda}(i)}.
\end{equation}
where $\widehat{\lambda}(i)$ is the eigenvalue of $\bm{M}$ with the biggest real part such that the $i$-th entry of at least one of its eigenvectors $\vec{c}_{p}$ is different from zero. 
Similarly, $\widehat{p}(i)$ is the maximum value of $p$
among these eigenvectors with non-zero $i$-th entries. 
In general, then, $\widehat{\lambda}(i)$ might not be the maximum eigenvalue of $\bm{M}$, like $\widehat{p}(i)$ might be less than the multiplicity of the eigenvalue $\widehat{\lambda}(i)$ minus one. Moreover, different nodes may have different values for these exponents. In particular, we have the same exponents for nodes in the same strongly connected components (SCCs), while they may vary from SCC to SCC. In the following paragraphs we will investigate this aspect.
%

\medskip

\textbf{\textit{Strongly connected network}}\textemdash
Let us suppose that the graph $G(\mathcal{N},\mathcal{E})$ is strongly connected. In this case the solution given by Eq.~\eqref{eq:approx_leading_solution_SM} simplifies. Indeed, in this case, the corresponding adjacency matrix $\bm{A} = \{a_{ij}\}$ is irreducible \cite{berman1974.ch2}. 
Let us recall that for irreducible matrices the Perron–Frobenius theorem holds \cite{perron1907uber, frobenius1912matrizen}, according to which there exists a positive eigenvalue $\widehat{\mu}$ greater or equal to (in absolute value) all other eigenvalues. Such eigenvalue corresponds to a simple root of the characteristic equation and the corresponding eigenvector $\vec{u}$ has all positive entries too. The latter vector is a multiple of the Bonacich eigenvector centrality vector \cite{bonacich1972factoring}. Widely used in network science, the Bonacich eigenvector centrality is a measure that recursively accounts for local and global properties of the network, relying on the notion that a node can be highly central either by having a high degree or by being connected to others that themselves are highly central~\cite{latora_nicosia_russo_2017}.
Simple algebraic steps can prove that if $\mu$ is an eigenvalue for $\bm{A}$, then $\lambda=f(\mu)$ is an eigenvalue for $\bm{M}$. Moreover, if $\vec{u}$ is an eigenvector corresponding to the eigenvalue $\mu$ of $\bm{A}$, then $\vec{u}$ is also an eigenvector corresponding to the eigenvalue $\lambda=f(\mu)$ of $\bm{M}$.
Therefore, if $\widehat{\mu}$ is the maximum eigenvalue of $\bm{A}$, then 
$\widehat{\lambda} = f(\widehat{\mu}) = \frac{\nu}{\rho} + \frac{\nu+1}{\rho} \widehat{\mu}>0$ is the highest eigenvalue of $\bm{M}$, and with the same positive eigenvector $\vec{u}$.
Thus, for strongly connected graphs, the approximated solution given by Eq.~\eqref{eq:approx_leading_solution_SM} becomes
\begin{equation}\label{eq:approx_connected_solution_suppl}
	D_i(t) \underset{t\to\infty}{\approx} u_i \,t^{\widehat{\lambda}},
\end{equation}
meaning that all nodes have similar Heaps' laws, and the key difference is made by their eigenvector centrality. As we saw in the main text (and we will see here more in details), these differences, more pronounced in transient times, will contribute to determine the fastest explorers in the network.
Moreover, we deduce that the approximation used in Eq.~\eqref{eq:approx_heaps_general_SM} is valid provided that $\widehat{\lambda} =  f(\widehat{\mu}) < 1$, that is $\rho > \nu + (\nu+1)\widehat{\mu}$, while for higher values of $\rho$ the solution is bounded by the linear solution as seen for the individual urn in Eq.~\eqref{eq:sol_isolated_urn}, since in the original system in Eq.~\eqref{eq:heaps_general_SM} we have $d_t D_i(t) \leq 1$.

\medskip
\textbf{\textit{Non-strongly connected network}}\textemdash
Let us now consider the most general case, that is a directed or undirected graph with any hypotheses of connectivity. Let us construct an algorithm to determine the pace of discovery of each node, which will help us better understand analytically why some nodes have higher paces of discovery. To do this, let us partition the graphs into its strongly connected components (SCCs), i.e. maximal strongly connected subgraphs of $G$, which can be found in linear computational time, for example with a DFS-based algorithm \cite{tarjan1972depth}. Let all the SCCs be indexed as $C_1, \dots, C_p$, with $C_i \cap C_j = \varnothing\; \forall i \neq j$. 

Without loss of generality, let us suppose that the graph $G$ is weakly connected, because otherwise we can repeat the same reasoning for each weakly connected component. Let us also suppose that the number of SCCs is $p>1$, because otherwise the graph would be strongly connected, which we already discussed in the previous paragraph. Since $G$ is weakly connected, for each SCC $C_q$ there must exist another component $C_l$, with $l\neq q$, such that there are some links from $C_q$ to $C_l$ or viceversa. However, there cannot be links in both directions (from $C_q$ to $C_l$ and viceversa), because otherwise they would be a unique SCC. It is also easy to show that there is always a SCC without any outgoing links to other SCCs.
Eventually permutating the indexes of the SCCs, let us call $C_1,\dots, C_{p_1}$  all the components with no outer links. Then, for each $1 \leq q \leq p_1$, the respective system of differential equations for $D_i$, $i\in C_q$, does not depend on any outer variable $D_j$, $j \in C_l \neq C_q$. Therefore, we can consider $C_q$ as an independent strongly connected subgraph of $G$, for which the reasoning in last paragraph holds. The solution for these SCCs is then:
\begin{equation}\label{eq:approx_connected_solution_C1}
	D_i(t) \underset{t\to\infty}{\approx} \gamma_i^{(q)} \, t^{\widehat{\lambda}^{(q)}}\  \forall i \in C_q, 1 \leq q \leq p_1,
\end{equation}
where $\widehat{\lambda}^{(q)}$ is the maximum eigenvalue of the adjacency matrix of subgraph $C_q$ and $\gamma_i^{(q)}$ is a multiple of the eigenvector centralitiy for node $i$ in $C_q$.
Found all the Heaps' laws relative to the nodes in $C_1, \dots, C_{p_1}$, it is possible to show that there exist SCCs $C_{p_1 + 1}, \dots, C_{p_2}$ that have links only towards the previously studied SCCs $C_1, \dots, C_{p_1}$, with $p_2 > p_1$. Then, choosing $C_q$ one of these other SCCs, let $\overline{\lambda}^{(q)}$ be the highest eigenvalue of the adjacency matrix of $C_q$. Let also $\tilde{\lambda}^{(q)} = \max_{l \leq p_1}(\gamma_{ql}\widehat{\lambda}^{(l)})$ be the maximum of the Heaps' exponents in Eq.~\eqref{eq:approx_connected_solution_C1} of the SCCs reachable from $C_q$, where $\gamma_{qr}=1$ if there is at least a link from $C_q$ to $C_l$, $\gamma_{qr}=0$ otherwise. As we will see further in this section, the Heaps' solutions for the nodes in these SCCs is:
\begin{equation}\label{eq:approx_connected_solution_other_SCC}
	D_i(t) \underset{t\to\infty}{\approx}
	\begin{cases}
		\gamma_i^{(q)} \,t^{\overline{\lambda}^{(q)}}\qquad  &\text{if } \overline{\lambda}^{(q)} > \tilde{\lambda}^{(q)} \\
		\gamma_i^{(q)} \ln(t)\,t^{\tilde{\lambda}^{(q)}}\qquad &\text{if } \overline{\lambda}^{(q)} = \tilde{\lambda}^{(q)} \\
		\gamma_i^{(q)} \,t^{\tilde{\lambda}^{(q)}}\qquad &\text{if } \overline{\lambda}^{(q)} < \tilde{\lambda}^{(q)}
	\end{cases}
	\quad\forall i \in C_q,\ p_1+1 \leq q \leq p_2,
\end{equation}
meaning that the Heaps' exponent $\widehat{\lambda}^{(q)}$ for node $i$ in $C_q$, $p_1+1 \leq q \leq p_2$, is 
\begin{equation}
	\widehat{\lambda}^{(q)} = \max (\overline{\lambda}^{(q)},\ \tilde{\lambda}^{(q)} ) ,
\end{equation}
that is the maximum of the highest eigenvalue $\overline{\lambda}^{(q)}$ of $\bm{M}$ relative to $C_q$ and the highest $\tilde{\lambda}^{(q)}$ of the Heaps' exponents $\widehat{\lambda}^{(l)}$ for $1\leq l\leq p_1$. Moreover, if $\overline{\lambda}^{(q)} = \tilde{\lambda}^{(q)}$, a factor $\ln(t)$ appears in the solution.
The same procedure can be repeated for all other successive SCCs $C_q$, keeping in mind that now a higher power $\ln^{\widehat{p}(q)}(t)$ of $\log(t)$ can appear. 

In this algorithmic process, let us now consider a generic SCC, say $C_q$, and let us suppose we have solved inductively all the equations for the Heaps' law of the nodes in the already examined SCCs, that is $C_1, \dots, C_{q-1}$. Let us recall that we arranged the indexes in such a way that the only outgoing links from $C_q$ are pointed to nodes in previous SCCs, i.e. in some of the SCCs $C_1, \dots, C_{q-1}$. For this reason, in order to solve the asymptotic differential equations responsible for the Heaps' law of the nodes in $C_q$, we can consider only the equations relative to the nodes in $C_q$ in Eq.~\eqref{eq:approx_heaps_general_matrix_SM}, since the previous SCCs have been already solved and the following variables do not appear in these equations. We hence have to solve the following approximated equations:
\begin{equation}\label{eq:approx_equation_C_q}
	\frac{dD_i(t)}{dt}\approx \frac{1}{t}\left(\frac{\nu}{\rho}D_i(t)+\frac{\nu+1}{\rho}\sum_{j\in C_q} a_{ij} D_j(t) +\frac{\nu+1}{\rho}\sum_{j\notin C_q} a_{ij} D_j(t)  \right), \quad i \in C_q,
\end{equation}
where we have isolated the contributions coming from nodes outside $C_q$, which we have suppose to be known.
Considering the general asymptotic solution for each individual Heaps' law derived for a strongly connected graph in Eq.~\eqref{eq:approx_leading_solution_SM}, for each $i \in C_q$ for large $t$ we can write explicitly the functions $D_j(t)$, $j\notin C_q$, which lets us write:
\begin{equation}\label{eq:approx_other_terms}
	\frac{\nu+1}{\rho}\sum_{j\notin C_q} a_{ij} D_j(t) \approx \frac{\nu+1}{\rho}\sum_{j\notin C_q} a_{ij}u_j \ln^{\widehat{p}_j}(t)\,t^{\widehat{\lambda}_j}
	\underset{t\to\infty}{\approx} \tilde{u}_i \ln^{\tilde{p}^{(q)}}(t)\,t^{\tilde{\lambda}^{(q)}} = f_i(t),
\end{equation}
where we have used the fact that  $\eta_i \ln^{\tilde{p}_i}(t)\, t^{\tilde{\lambda}_i}$ is the leading term of the expression $\sum_{j\notin C_q} a_{ij}u_j\ln^{\widehat{p}_j}(t)\,t^{\widehat{\lambda}_j}$ and that we are working for large $t$. %
Then, using Eq.~\eqref{eq:approx_other_terms} and calling $\vec{D}^{(q)}$ and $\bm{A}^{(q)}$ the sub-vector of $\vec{D}$ and sub-matrix of $\bm{M}$ relative to $C_q$, we can rewrite Eq.~\eqref{eq:approx_equation_C_q} in a compact form as
\begin{equation}\label{eq:approx_system_C_q}
	\frac{d\vec{D}^{(q)}(t)}{dt}\approx 
	\frac{\bm{M}^{(q)} \vec{D}^{(q)}(t)}{t} + \frac{\vec{f}^{(q)}(t)}{t}.
\end{equation}
The associated homogeneous system corresponds to the considering the sub-graph $C_q$ without all the external links. For this system we get the same solution derived for a strongly connected graph  in Eq.~\eqref{eq:approx_connected_solution_suppl}, which is 
\begin{equation}\label{eq:sol_hom_C_q}
	\vec{\overline{D}}^{(q)}(t) \underset{t\to\infty}{\approx}
	\vec{\overline{u}}^{(q)} \,t^{\overline{\lambda}^{(q)}},
\end{equation}
where $\overline{\lambda}^{(q)}$ is the highest eigenvalue of $\bm{M}^{(q)}$ (positive and simple for the Perron-Frobenious theorem), and $\vec{u}^{(q)}$ is a multiple of the eigenvector centrality. Let us search a solution for Eq.~\eqref{eq:approx_system_C_q} of the form 
$\vec{D}^{(q)}(t) = \vec{u}^{(q)}(t) \circ \vec{\overline{D}}^{(q)}(t)$, where  $\circ$ is the Hadamard (element-wise) product, 
that plugged in Eq.~\eqref{eq:approx_system_C_q} gives:
\begin{equation}\label{eq:approx_system_C_q_2}
	\frac{d\vec{u}^{(q)}(t)}{dt}\circ \vec{\overline{D}}^{(q)}(t) + \cancel{\vec{u}^{(q)}(t) \circ \frac{d\big[\vec{\overline{D}}^{(q)}(t)\big]}{dt}} \approx \cancel{\vec{u}^{(q)}(t)\circ \frac{\bm{M}^{(q)}  \, \vec{\overline{D}}^{(q)}(t)}{t}} + \frac{\vec{f}^{(q)}(t)}{t},
\end{equation}
where the cancellation is due to the general solution in Eq.~\eqref{eq:sol_hom_C_q} of the associated homogeneous system. Therefore, recalling Eq.~\eqref{eq:approx_other_terms} and Eq.~\eqref{eq:sol_hom_C_q} we have:
\begin{equation}\label{eq:approx_system_C_q_3}
	\frac{d \vec{u}^{(q)}(t)}{dt}  \approx  \vec{\tilde{u}}^{(q)} \circ \left(\vec{\overline{u}}^{(q)}\right)^{-1} \frac{\ln^{\tilde{p}^{(q)}}(t)\, t^{\tilde{\lambda}^{(q)}}} {t^{\overline{\lambda} + 1}} = \vec{\gamma}\, \frac{\ln^{\tilde{p}^{(q)}}(t)\, t^{\tilde{\lambda}^{(q)}}} {t^{\overline{\lambda} + 1}}, 
\end{equation}
or equivalently, considering the $i$-th components:
\begin{equation}\label{eq:approx_system_C_q_4}
	\frac{d u_i(t)}{dt}  \approx  \tilde{u}_i^{(q)} {\left[\left(\vec{\overline{u}}^{(q)}\right)^{-1}\right]}_i \frac{\ln^{\tilde{p}^{(q)}}(t)\, t^{\tilde{\lambda}^{(q)}}} {t^{\overline{\lambda} + 1}} = \gamma_i \frac{\ln^{\tilde{p}^{(q)}}(t)\, t^{\tilde{\lambda}^{(q)}}} {t^{\overline{\lambda} + 1}},  
\end{equation}
where we have defined  $\vec{\gamma} = \vec{\tilde{u}}^{(q)} \circ \left(\vec{\overline{u}}^{(q)}\right)^{-1}$   
and $\gamma_i = \tilde{u}_i^{(q)} {\left[\left(\vec{\overline{u}}^{(q)}\right)^{-1}\right]}_i$ its $i$-th component.
Let us hence distinguish three cases. 
\begin{enumerate}
	\item If $\overline{\lambda}^{(q)} > \tilde{\lambda}^{(q)}$, then we have:
	\begin{equation}\label{eq:sol_c1_terzo_C_q}
		u_i(t) \approx \frac{\gamma_i}{\tilde{\lambda}^{(q)}-\overline{\lambda}^{(q)}} \ln^{\tilde{p}^{(q)}}(t)\,t^{\tilde{\lambda}^{(q)}-\overline{\lambda}^{(q)}} + u_i \underset{t\to\infty}{\approx} u_i,
	\end{equation}
	which gives the solution:
	\begin{equation}\label{eq:sol_S_terzo_C_q}
		D_i(t) \underset{t\to\infty}{\approx} u_i t^{\overline{\lambda}^{(q)}}.
	\end{equation}
	
	\item Similarly, for $\overline{\lambda}^{(q)} = \tilde{\lambda}^{(q)}$ we have:
	\begin{equation}\label{eq:sol_c1_secondo_C_q}
		u_i(t) \approx \frac{\gamma_i}{\tilde{p}^{(q)}+1} \ln^{\tilde{p}^{(q)}+1}(t) + u_i \underset{t\to\infty}{\approx} \frac{\gamma_i}{\tilde{p}^{(q)}+1} \ln^{\tilde{p}^{(q)}+1}(t),
	\end{equation}
	which gives:
	\begin{equation}\label{eq:sol_S_secondo_C_q}
		D_i(t) \underset{t\to\infty}{\approx} u_i  \ln^{\tilde{p}^{(q)}+1}(t)\,t^{\tilde{\lambda}^{(q)}}.
	\end{equation}
	
	\item Finally, if $\overline{\lambda}^{(q)} < \tilde{\lambda}^{(q)}$ we have:
	\begin{equation}\label{eq:sol_c1_primo_C_q}
		u_i(t) \approx \frac{\gamma_i}{\tilde{\lambda}^{(q)}-\overline{\lambda}^{(q)}} \ln^{\tilde{p}^{(q)}}(t)\,t^{\tilde{\lambda}^{(q)}-\overline{\lambda}^{(q)}} + d_1 \underset{t\to\infty}{\approx} \frac{\gamma_i}{\tilde{\lambda}^{(q)}-\overline{\lambda}^{(q)}} \ln^{\tilde{p}^{(q)}}(t)\,t^{\tilde{\lambda}^{(q)}-\overline{\lambda}^{(q)}},
	\end{equation}
	hence the solution:
	\begin{equation}\label{eq:sol_S_primo_C_q}
		D_i(t) \underset{t\to\infty}{\approx} a_i \ln^{\tilde{p}^{(q)}}(t)\,t^{\tilde{\lambda}^{(q)}}.
	\end{equation}
\end{enumerate}

To sum up, we have the following solutions:
\begin{equation}\label{eq:algorithm_solution_general}
	D_i(t) \underset{t\to\infty}{\approx}
	\begin{cases}
		u_i t^{\overline{\lambda}^{(q)}} \qquad  &\text{if } \overline{\lambda}^{(q)} > \tilde{\lambda}^{(q)} \\
		u_i \ln^{\tilde{p}^{(q)}+1}(t)\, t^{\tilde{\lambda}^{(q)}}\qquad &\text{if } \overline{\lambda}^{(q)} = \tilde{\lambda}^{(q)} \\
		u_i\ln^{\tilde{p}^{(q)}}(t)\, \,t^{\tilde{\lambda}^{(q)}}\qquad &\text{if } \overline{\lambda}^{(q)} < \tilde{\lambda}^{(q)}
	\end{cases}
	\quad\forall i \in C_q, q > p_1,
\end{equation}
Comparing this solution with the general one we gave in Eq.~\eqref{eq:approx_leading_solution_SM}, we have (a)~$\widehat{\lambda}(i) = \overline{\lambda}^{(q)}$ and $\widehat{p}(i) = 0$ if $\overline{\lambda} > \tilde{\lambda}$, (b)~$\widehat{\lambda}(i) = \tilde{\lambda}^{(q)}$ and $\widehat{p}(i) = \tilde{p}^{(q)} + 1$ if $\overline{\lambda} = \tilde{\lambda}$,  and  (c)~$\widehat{\lambda}(i) = \tilde{\lambda}^{(q)}$ and $\widehat{p}(i) = \tilde{p}^{(q)}$ if $\overline{\lambda} < \tilde{\lambda}$. 

In conclusion, when dealing with a network with multiple strength connected components, we solve the equations for the components that are independent from the others. Then we consider the SCCs that have links only to previous SCCs, applying the method just described. This is repeated until every SCC is studied, thus solving the whole system and describing the pace of discovery of each node of the entire network analytically, obtaining solutions of the type in Eq.~\eqref{eq:approx_leading_solution_SM}. In the next section this algorithmic method is applied to simple networks with $N=4$ nodes, as we have already implicitly done above for a two nodes network and for chains.

\subsection{Application to the five graphs in Fig.~3}\label{sec:approx_Heaps_other_examples}

As an application of the analytical results of the previous sections, we study here the very same five networks reported in Fig.~3 of the main text. In particular, we will be able to provide an explicit expression for the growth of the number of novelties 
at each of the four nodes of the social network.\medskip

\textbf{\textit{Graph a}}\textemdash
Let us consider a network where nodes 2, 3, and 4 do not have any outgoing links, while node 1 has the links $1\rightarrow2$, $1\rightarrow3$, and $1\rightarrow4$ to all other nodes (see network representation in Table~\ref{tab:explicit_solutions_little_graphs}).
Let us observe that the dynamics here is very similar to the case of a couple of urns with the only link $1\rightarrow2$. 
Nodes 2, 3, and 4 can be considered as three individual urns, for which the Heaps' law is the same to the classic one in Eq.~\eqref{eq:isolated_urn}, that is:
\begin{equation}\label{eq:isolated_urn_a}
	D_2(t) \underset{t\to\infty}{\approx} D_3(t) \underset{t\to\infty}{\approx} D_4(t) \underset{t\to\infty}{\approx} (\rho-\nu)^{\frac{\nu}{\rho}}t^{\frac{\nu}{\rho}}.
\end{equation}
As for node 1, the differential equation for the Heaps' law is approximated by:
\begin{equation}\label{eq:heaps_approx_a}
	\frac{dD_1(t)}{dt} \approx \frac{\nu D_1(t)}{\rho t} + \frac{(\nu+1)\big(D_2(t)+D_3(t)+D_4(t)\big)}
	{\rho t} \approx
	\frac{\nu D_1(t)}{\rho t} + \frac{3(\nu+1)(\rho-\nu)^{\frac{\nu}{\rho}}t^{\frac{\nu}{\rho}}}
	{\rho t}.
\end{equation}
The resolution of Eq.~\eqref{eq:heaps_approx_a} is the same as the one done for the couple of urns, with only a multiplicative factor 3. Therefore, the Heaps' solution for node 1 is:
\begin{equation}\label{eq:solution_a1}
	D_1(t) \underset{t\to\infty}{\approx} 3\, \frac{\nu+1}{\rho} (\rho-\nu)^{\frac{\nu}{\rho}}\ln(t)\ t^{\frac{\nu}{\rho}},
\end{equation}
which means that node 1 has a higher pace of discovery than nodes 2, 3, and 4, but at asymptotic times they will show the same Heaps' exponent. Moreover, it is clear that in star-like networks adding more nodes does not increase significantly the pace of discovery.\medskip

\textbf{\textit{Graph b}}\textemdash
The next network we studied is a chain of 4 nodes, with links $1\rightarrow2$, $2\rightarrow3$, and $3\rightarrow4$. This network has already been studied in Sec.~1.3, and the solutions are:
\begin{equation}
	D_{i}(t)\approx\frac{(\rho-\nu)^{\nu/\rho}}{(4-i)!}\left(\frac{\nu+1}{\rho}\ln(t)\right)^{4-i}t^{\nu/\rho}, \quad i = 1,2,3,4.
\end{equation}
This analytical result shows us why node 1 has an higher pace of discovery than the other nodes, with lower Heaps' exponents for higher nodes. This is due to the presence of different powers of the logarithm. In the end, however, they all have the same asymptotic Heaps' exponent, meaning that the difference is visible only at finite times.\medskip

\textbf{\textit{Graph c}}\textemdash
Let us consider a network made by a directed cycle between nodes 2, 3 and 4, with links $2\rightarrow3$, $3\rightarrow4$, and $4\rightarrow2$, and another node 1 linked directly to node 2 ($1 \rightarrow 2$).
In this case, we can distinguish two SCCs, the cycle and node 1. Since there is no link going out from the cycle, we start solving the Heaps' law equations related to it. As we have seen in Sec.~1.4, the solution is given by Eq.~\eqref{eq:sol_cycle_directed} with N = 3, that is:
\begin{equation}\label{eq:solution_c_cycle}
	D_i(t) \underset{t\to\infty}{\approx} (\rho-2\nu-1)^{\frac{2\nu+1}{\rho}}t^{\frac{2\nu+1}{\rho}}, \quad i = 2,3,4.
\end{equation}
Now let us consider the remaining SCC, namely node 1. Its equation is the same as Eq.~\eqref{eq:heaps_coupled_urn_explicit} for the two coupled urns case in Sec.~1.2, with the only difference that here the solution of $D_2(t)$ has a higher exponent. Then, if we search for a solution like $D_1(t) = \kappa(t)\overline{D}_1(t)$, with $\overline{D}_1(t) \approx (\rho-\nu)^{\frac{\nu}{\rho}}t^{\frac{\nu}{\rho}}$ 
being the solution of the associated homogeneous equation, we get:
\begin{equation}
	\frac{d\kappa(t)}{dt}=\frac{\nu+1}{\rho t}\frac{D_2(t)}{\overline{D}_1(t)}\approx\frac{\nu+1}{\rho t} \frac{(\rho-2\nu-1)^{\frac{2\nu+1}{\rho}}t^{\frac{2\nu+1}{\rho}}}{(\rho-\nu)^{\frac{\nu}{\rho}}t^{\frac{\nu}{\rho}}} = 
	\frac{\nu+1}{\rho} \frac{(\rho-2\nu-1)^{\frac{2\nu+1}{\rho}}t^{\frac{\nu+1}{\rho}-1}}{(\rho-\nu)^{\frac{\nu}{\rho}}},
\end{equation}
whose solution is:
\begin{equation}\label{eq:k_t_example_1}
	\kappa(t) \approx \frac{\nu+1}{\rho} \frac{(\rho-2\nu-1)^{\frac{2\nu+1}{\rho}}t^{\frac{\nu+1}{\rho}}}{(\rho-\nu)^{\frac{\nu}{\rho}}},
\end{equation}
which gives the asymptotic solution:
\begin{equation}
	D_1(t)\approx \frac{\nu+1}{\rho} (\rho-2\nu-1)^{\frac{2\nu+1}{\rho}} t^{\frac{2\nu+1}{\rho}}.
\end{equation}
We could have obtained the same result using the algorithm developed in the last section. In this case, node 1 gets the same dynamics of the nodes in the cycle, with just a scaling factor $(\nu+1)/\rho$, since the maximum eigenvalue of its SCC (node 1 itself) is lower than the maximum eigenvalue of the SCCs he is linked to (the cycle).\medskip

\textbf{\textit{Graph d}}\textemdash
In this case we consider the same network as the last graph we just analyzed swapping the direction of the link $4\rightarrow2$. Therefore, the cycle is broken (see network representation in Table~\ref{tab:explicit_solutions_little_graphs}), and as we are about to see, the dynamics is much more similar to a chain. We could give a detailed solution as done for the chain; instead, we are going to use directly the algorithm we developed to assess all the exponents in the Heaps' laws for every node.
Let us start from node 4, which has no outgoing links. This node is hence an individual urn, with the usual solution:
\begin{equation}\label{eq:heaps_d_4}
	D_4(t) \underset{t\to\infty}{\approx} (\rho-\nu)^{\frac{\nu}{\rho}}t^{\frac{\nu}{\rho}}.
\end{equation}
Let us move on to the SCC with outgoing links only towards previously studied SCCs, that is the SCC composed by node 3. If this SCC had no outgoing links, then it would be an isolated urn, therefore with the same exponent of the other SCC studied (node 4), meaning that the actual solution for node 3 has that exponent and a logarithmic factor. Indeed, the dynamics of node 3 is the same derived for the couple of urns in Sec.~1.2, which is:
\begin{equation}\label{eq:heaps_d_3}
	D_3(t) \underset{t\to\infty}{\approx} \frac{\nu+1}{\rho}(\rho-\nu)^{\frac{\nu}{\rho}}\, \ln(t)\,t^{\frac{\nu}{\rho}}.
\end{equation}
Proceeding with node 2, we compare its exponent if it was isolated to the maximum of the exponents of node 3 and 4, which are all the same. Moreover, since node 3 has a higher power in the logarithm than node 4, in the asymptotic solution, we can disregard the presence of the link $4\rightarrow2$. Thus, the solution for node 2 has another logarithmic factor and another constant multiplicative factor than those of node 3, that is we have the solution:
\begin{equation}\label{eq:heaps_d_2}
	D_2(t) \underset{t\to\infty}{\approx} \left(\frac{\nu+1}{\rho}\right)^{2}(\rho-\nu)^{\frac{\nu}{\rho}}\, \ln^2(t)\,t^{\frac{\nu}{\rho}}.
\end{equation}
To complete, similarly we obtain the solution for node 1, i.e.:
\begin{equation}\label{eq:heaps_d_1}
	D_1(t) \underset{t\to\infty}{\approx} \left(\frac{\nu+1}{\rho}\right)^{3}(\rho-\nu)^{\frac{\nu}{\rho}}\, \ln^3(t)\,t^{\frac{\nu}{\rho}}.
\end{equation}
We can hence see that the solutions are equal to those of the chain in Sec.~1.7.b, and there are only some slight differences at finite times due to the presence of another link, but not significantly.\medskip

\begin{table}[t]
	\centering 
	\begingroup
	\setlength{\tabcolsep}{12pt} 
	\renewcommand{\arraystretch}{1.5} 
	\begin{tabular}{|c| ccccc|}
		\toprule 
		& 	\raisebox{-.2\height}{\includegraphics[width=1.7cm]{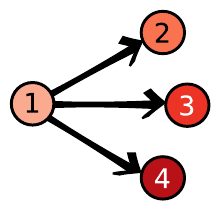}} & 
		\raisebox{-.2\height}{\includegraphics[width=1.7cm]{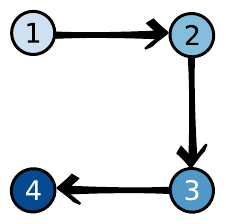}} &
		\raisebox{-.2\height}{\includegraphics[width=1.7cm]{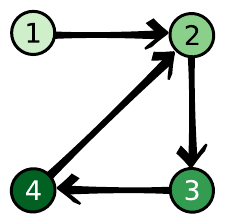}} &
		\raisebox{-.2\height}{\includegraphics[width=1.7cm]{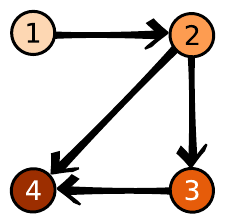}} &
		\raisebox{-.2\height}{\includegraphics[width=1.7cm]{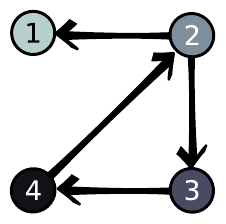}} \\
		\textbf{Network} & (a) & (b) & (c) & (d) & (e) \\
		\toprule 
		$D_1(t)\underset{t\to\infty}{\approx}$ & $u_1\ln(t)\,t^{\frac{\nu}{\rho}}$ & $u_1\ln^3(t)\,t^{\frac{\nu}{\rho}}$ & $u_1\,t^{\frac{2\nu+1}{\rho}}$ & $u_1\ln^3(t)\,t^{\frac{\nu}{\rho}}$ & $u_1\,t^{\frac{\nu}{\rho}}$ \\
		$D_2(t)\underset{t\to\infty}{\approx}$ & $u_2\,t^{\frac{\nu}{\rho}}$ & $u_2\ln^2(t)\,t^{\frac{\nu}{\rho}}$ & $u_2\,t^{\frac{2\nu+1}{\rho}}$ & $u_2\ln^2(t)\,t^{\frac{\nu}{\rho}}$ & $u_2\,t^{\frac{2\nu+1}{\rho}}$ \\
		$D_3(t)\underset{t\to\infty}{\approx}$ & $u_3\,t^{\frac{\nu}{\rho}}$ & $u_3\ln(t)\,t^{\frac{\nu}{\rho}}$ & $u_3\,t^{\frac{2\nu+1}{\rho}}$ & $u_3\ln(t)\,t^{\frac{\nu}{\rho}}$ & $u_3\,t^{\frac{2\nu+1}{\rho}}$ \\
		$D_4(t)\underset{t\to\infty}{\approx}$ & $u_4\,t^{\frac{\nu}{\rho}}$ & $u_4\,t^{\frac{\nu}{\rho}}$ & $u_4\,t^{\frac{2\nu+1}{\rho}}$ & $u_4\,t^{\frac{\nu}{\rho}}$ & $u_4\,t^{\frac{2\nu+1}{\rho}}$ \\
		\toprule 
	\end{tabular}
	\endgroup
	\caption{Summary of the asymptotic Heaps' laws derived analytically for the 4 nodes composing the five networks reported in Fig.~3 of the main text (here displayed at the top). The coefficients $u_i$ have not been reported to focus on the exponents of the power laws and the logarithms, when present. }
	\label{tab:explicit_solutions_little_graphs}
\end{table}

\textbf{\textit{Graph e}}\textemdash
The last case to be examined is again similar to Graph~c, but this time we swap the direction of the link between nodes 1 and 2 (see network representation in Table~\ref{tab:explicit_solutions_little_graphs}). Here the order with which we study the SCCs is inverted, because now only node 1 has no outer links. Therefore, the Heaps' law for node 1 is the classic individual one in Eq.~\eqref{eq:sol_isolated_urn}. Then we have to solve the equations for the cycle, which in this case are:
\begin{equation}\label{eq:heaps_eq_e}
	\begin{cases}
		\dfrac{dD_2(t)}{dt} \approx
		\dfrac{\nu D_2(t)}{\rho t} + \dfrac{(\nu+1)D_3(t)}{\rho t} + \dfrac{(\nu+1)D_1(t)}{\rho t} \\[0.3cm]
		\dfrac{dD_3(t)}{dt} \approx 
		\dfrac{\nu D_3(t)}{\rho t} + \dfrac{(\nu+1)D_4(t)}{\rho t} \\[0.3cm]
		\dfrac{dD_4(t)}{dt} \approx 
		\dfrac{\nu D_4(t)}{\rho t} + \dfrac{(\nu+1)D_2(t)}{\rho t} .
	\end{cases}
\end{equation}
In this system, we can consider $D_1(t)$ known, working at large time-scales. Therefore, following the algorithm described in Sec.~1.6.2, we first solve this system without the external sources (i.e. node 1), in order to find the leading solution and then compare the exponents with the external sources ones. The solution of the associated homogeneous system is the same of a directed cycle as in Eq.~\eqref{eq:sol_cycle_directed}, i.e. a power-law function with exponent $2\nu+1/\rho$. Now, we observe that the Heaps' exponent of the cycle is higher than the exponents of outer SCCs it is linked to, that is just node 1 with exponent $\nu/\rho$. Then, the asymptotic solution for the nodes in the cycle corresponds to the solution of the cycle as if it had no outer links. Explicit solutions are given in Table~\ref{tab:explicit_solutions_little_graphs}.


\section{Node ranking and Heaps'law}\label{sec:heaps_laws_ranking}

In this section, we study more in details the validity of the eigenvector centrality and $\alpha$-centrality to rank the nodes in a social network according to their discovery dynamics. First, we describe the real-world data sets considered. Then, we test the persistence of the nodes ranking based on the fitted Heaps' exponents at different times. Finally, we explain why the eigenvector centrality and the $\alpha$-centrality lead to the same ranking of the Heaps' exponents for strongly-connected and generic networks respectively. All simulations in this section are performed with model parameters: $\rho=10$, $\nu=1$, $M_0=\nu+1$.


\subsection{Description of the data sets}
We consider four data sets of real-world networks representing different types of social interactions: the Zachary Karate Club (ZKC) network~\cite{zachary1977information}, a network of follower relationships among Twitter users~\cite{de2010does}, a co-authorship network in Network Science~\cite{newman2006finding}, and a collaboration network between jazz musicians~\cite{gleiser2003community}. The network of Twitter from the original data set (Ref.~\cite{de2010does}) has been reduced by performing a random walk sampling.

Some basic properties of the networks are summarized in Table~\ref{table:real_world_nets}, like the total number of nodes $N$, the total number of links $E$, the average degree $\langle k\rangle$, and the maximum eigenvalue~$\widehat{\mu}$ of the related adjacency matrix. Moreover, we have shown some properties of connection of the networks. In particular, we distinguished weakly-connected components (CCs) and strongly-connected components (SCCs), because they play an important role in the dynamics under investigation. Therefore we showed the number of both CCs and SCCs, as well as the size of the respective largest one. As we can see, the networks we have chosen have all very different properties, either in size, average degree, and connection. 
\begin{table}[htbp]
	\begin{center} 
		\begin{tabular}{|l| c | c | c | c | c | c | c | c | c | c |} 
			\hline
			{\bf Data set} & Label & Type & $N$ & $E$ & $\langle k\rangle$ & $\widehat{\mu}$ & Num. CCs & Num. SCCs & Size LCC & Size LSCC\\
			\hline
			ZKC & (a) & Undirected & 34   & 78    &4.6 & 6.7 & 1 & 1 & 34 & 34\\
			Twitter & (b) & Directed & 4968 & 26875 & 10.8 & 5.2 & 1 & 4164 & 4968 & 770\\
			NetSci & (c) & Undirected & 1589 & 2742  &3.4 & 19.0 & 396 & 396 & 379 & 379\\
			Jazz & (d) & Undirected & 198  & 2742  &27.7 & 40.0 & 1 & 1 & 198 & 198\\
			\hline
		\end{tabular}
		\caption[]{Statistics and properties of the four real-world networks considered (see Fig.~4 of the main text): number of nodes $N$, number of edges $E$, average node degree $k$, maximum eigenvalue $\widehat{\mu}$, number of (weakly) connected components (CCs), and number of strongly connected components (SCCs), size of the largest (weakly) connected component (LCC), and size of the largest strongly connected component (LSCC).}
		\label{table:real_world_nets}
	\end{center}
\end{table}


\subsection{Rank persistence}

\begin{figure}
	\includegraphics[width=0.7\textwidth]{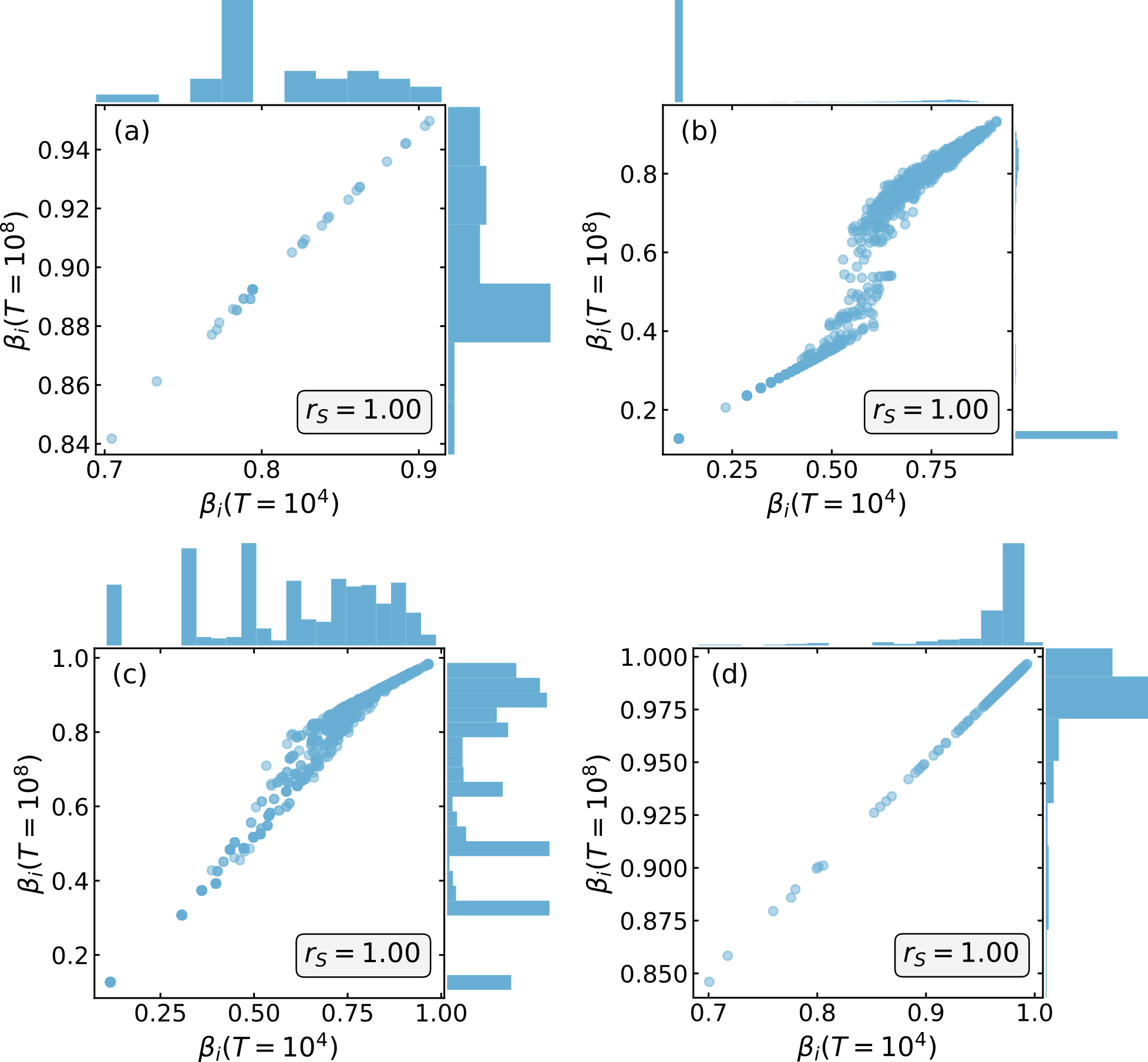}
	\caption{Scatter plot and Spearman’s rank correlation coefficient $r_{S}$ between fitted Heaps’ exponents $\beta_i(T)$ at $T=10^4$ and $T=10^8$ associated to the $i=1,\dots,N$ nodes off the four empirical networks considered: (a) the Zachary Karate Club network \cite{zachary1977information}, (b) a network of follower relationships of Twitter \cite{de2010does}, (c) a co-authorship network in network science \cite{newman2006finding} and (d) a collaboration network between jazz musicians \cite{gleiser2003community}. The parameters of the model are $\rho=10$, $\nu=1$, $M_0=\nu+1$.
		\label{fig:SI_spearman_betas}
	}
\end{figure}

In the main text, we have developed a networked model for the dynamics of discovery that introduces an heterogeneity in the paces of discovery, as it happens in real-world social networks. In the previous sections, we concentrated on finding an analytical asymptotic solution of the Heaps' laws. However, for most of applications we are interested in transient times. As can be seen in Fig.~3 of the main text, the paces of discoveries, represented by the fitted Heaps' exponents, change in time, depending on the network characteristics and the model parameters. Nonetheless, the ranking of the nodes based on these fitted exponents remains almost the same. To show this, we plot in Fig.~\ref{fig:SI_spearman_betas} the scatter plot and the Spearman's rank correlation coefficient between the fitted Heaps' exponents $\beta(T)$ at $T=10^4$ and $T=10^8$, together with their distributions, for the four real-world networks presented in the last section. In all cases, we get a Spearman's correlation of 1.00, meaning that even though the distribution of fitted exponents change, the ranking is time-invariant and does not depend on the particular $T$ at which Heaps' exponents are fitted.
Let us observe that we used a set of parameters that in all cases invalidate the approximations used in the analytical study, i.e. $\rho < \nu + (\nu+1)\widehat{\mu}$.

This is evident in the scatter plot of Fig.~\ref{fig:SI_spearman_betas}(b), where, apart from a set of nodes whose exponents span across the entire range, most of the nodes present a very low pace of discovery, with fitted exponents very close to 0. A similar thing can be seen in Fig.~\ref{fig:SI_spearman_betas}(d), for which we have the highest eigenvalue and hence the highest Heaps' exponents among the four networks (with all Heaps' exponents very close to 1).
All  this is a strong indication that the various paces of discovery have to depend on some structural characteristics of the networks.

In the following sections, we keep investigating the relations between Heaps' exponents and network measures. In particular, we focus on the eigenvector centrality and the $\alpha$-centrality, respectively useful for strongly-connected graphs and generic graphs. More insights on these centrality measures will be provided, both from a numerical and an analytical point of view.

\subsection{Heaps' exponents and the eigenvector centrality}
In the main text, we have shown that in strongly connected graphs each urn has the same asymptotic Heaps' exponent, and the driving factor for each node is the associated asymptotic coefficient. As we saw when we derived the asymptotic expression of the Heaps' law for strongly connected graphs in Eq.~\eqref{eq:approx_leading_solution_SM}, the Heaps' exponent corresponds to the maximum eigenvalue $\widehat{\lambda}$ of the matrix $\bm{M} = \frac{\nu}{\rho} \bm{I} + \frac{\nu+1}{\rho} \bm{A}$, where $\bm{A}$ is the adjacency matrix. In particular, because of the Perron-Frobenius theorem \cite{perron1907uber, frobenius1912matrizen}, we know that $\widehat{\lambda}$ is positive and simple, and the related eigenvector $\vec{u}$ has all positive entries. We also derived that the coefficients of the Heaps' laws are all multiples of this eigenvector. A lot of importance has been given in the past to this vector, from which we can derive the eigenvector centrality, also known as the Bonacich centrality~\cite{bonacich1972factoring}. As a definition, the eigenvector centrality $c_i^{(E)}$ of node $i$ is the $i$-th coefficient of the normalized solution of the equation:
\begin{equation}\label{eq:eigenvector_centrality}
	\bm{M}\vec{c}^{\,(E)}=\widehat{\lambda}\, \vec{c}^{\,(E)}, 
\end{equation}
where $\widehat{\lambda}$ is the highest positive eigenvalue~\cite{perron1907uber}. This centrality measure accounts for both local and global properties of the network, as it is not just dependent on the degree of the node, but also on the positioning of each node in the network~\cite{lu2016vital}. 

Our analytical investigation showed us that for strongly connected components we expect the same asymptotic Heaps' exponents. However, the same analysis showed us that the coefficients depend on the eigenvector centrality. This factor plays a role in the transient times, when we are far from the asymptotic regime, and it is thus especially important for real-world systems.

\begin{figure}
	\centering
	\includegraphics[width=0.8\linewidth]{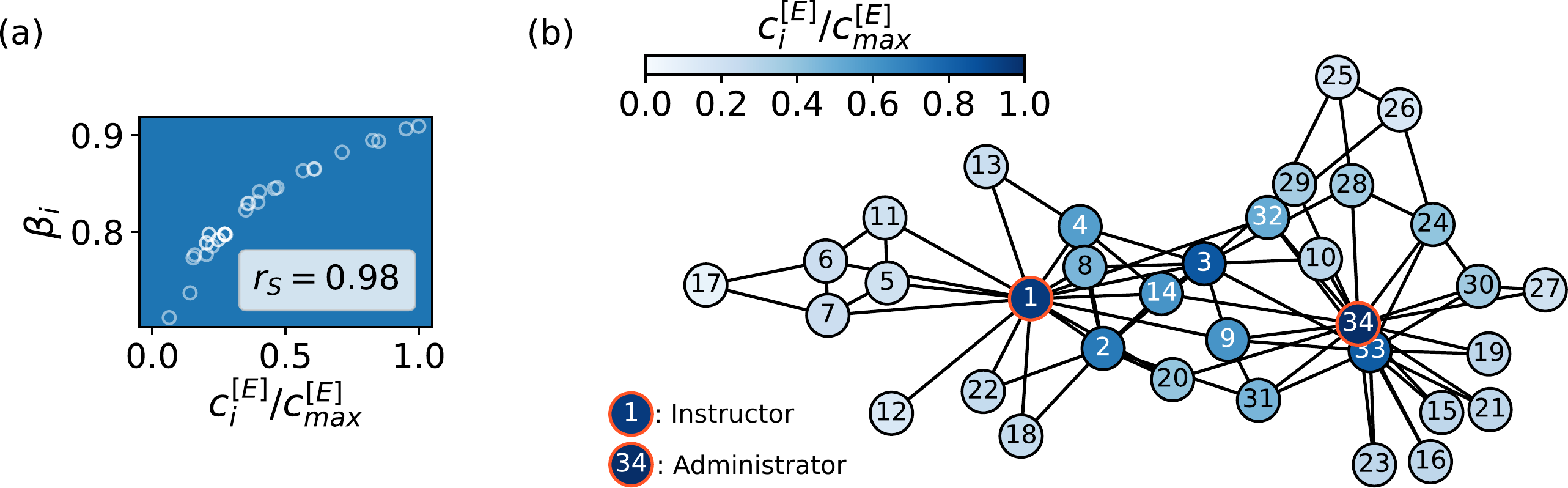}
	\caption{Dynamics of the interacting urns on the Zachary Karate Club network \cite{zachary1977information}. (a) Scatter plot and Spearman's rank correlation coefficients $r_{S}$ between fitted Heaps' exponents $\beta_i(T=10^4)$ and normalized eigenvector centrality $c^{[E]}_i/c^{[E]}_{\text{max}}$ associated to the $i=1,\dots,N$ nodes of the network. (b) Nodes are colored according to the resulting normalized eigenvector centrality.}
	\label{fig:SI_spearman_rank_corr_ZKC}
\end{figure}

To complement the results presented in the main text, we now test numerically the correlation between the eigenvector centralities and the measured Heaps' exponents at transient times for the Zachary Karate Club network.
Figure \ref{fig:SI_spearman_rank_corr_ZKC}(a) shows the scatter plot and the Spearman's rank correlation of the eigenvector centralities and the fitted Heaps' exponents at time $T=10^4$ for the (largest strongly connected component of) ZKC network and in Fig.~\ref{fig:SI_spearman_rank_corr_ZKC}(b) its visualization with color-coded nodes (cfr Fig.~2 of the main text).
The resulting Spearman's rank correlation higher than $0.98$ persists changing the parameters in the simulations, even for sets of parameters in contrast with the approximations used in the analytical study, i.e. $\rho < \nu + (\nu+1)\widehat{\mu}$. We can hence conclude that the eigenvector centrality is an optimal proxy for the distribution of Heaps' exponents in strongly connected social networks, and it can be used to give a faithful ranking of the individual expected paces of discovery. 

%

\subsection{Heaps' exponents and the \texorpdfstring{$\bm{\alpha}$}{alpha}-centrality}

In this section we focus on generic directed graphs and the usage of the $\alpha$-centrality as a proxy for the ranking of the nodes based on their pace of discovery in these more general cases.
The $\alpha$-centrality, widely used in network analysis \cite{Ide2014centrality,Ghosh2012centrality}, has been first introduced in Ref.~\cite{bonacich2001eigenvector} to extend the eigenvector centrality to asymmetric graphs. 
The underlying idea is to tune the influence of the adjacency matrix structure with a parameter $\alpha$ to add exogenous sources to the centrality~\cite{bonacich2001eigenvector, latora_nicosia_russo_2017}. Formally, it is defined as the vector $\vec{u}$ such that
\begin{equation}\label{eq:alpha_centrality_def_1}
	\vec{c}^{\,(\alpha)} = \alpha \bm{A} \vec{c}^{\,(\alpha)} + \vec{e},
\end{equation}
where $\vec{e}$ is an $N$-dimensional vector of ones. 
The matricial form of Eq.~\eqref{eq:alpha_centrality_def_1} reads:
\begin{equation}\label{eq:alpha_centrality_def_2}
	\vec{c}^{\,(\alpha)} = (\bm{I}-\alpha \bm{A})^{-1}\vec{e} = \left(\sum_{k=0}^\infty \alpha^k \bm{A}^k\right)\vec{e},
\end{equation}
where $\bm{I}$ is the $N$-dimensional identity matrix. 
It has also been shown that this centrality is equivalent to Katz-centrality \cite{katz1953} given by 
\begin{equation}
	\vec{c}^{\,(K)} = \left(\sum_{k=1}^\infty a^k\bm{A}^k\right)\vec{e},
\end{equation}
with $a$ being an attenuation factor. In fact, it has been shown that the equality $\vec{c}^{\,(K)} = -\vec{e} + \vec{c}^{\,(\alpha)}$ holds, i.e. these two centralities differ only by a constant \cite{bonacich2001eigenvector}.
From Eq.~\eqref{eq:alpha_centrality_def_1} and~\eqref{eq:alpha_centrality_def_2}, it is clear that the $\alpha$-centrality can be both a local and global measure. In fact, for $\alpha \to 0^+$, the relative importance of the structure given by the adjacency matrix $\bm{A}$ decreases, in favor of the exogenous factor given by $\vec{e}$. With higher values of $\alpha$, instead, the role of the exogenous part is damped. 

For an undirected graph, the $\alpha$-centrality becomes proportional to the eigenvector centrality when $\alpha\to (1/\widehat{\mu})^-$, where $\mu$ is the highest positive eigenvalue of the adjacency matrix. In fact, in this case all eigenvalues are real and the eigenvectors are orthogonal. 
Following Ref.~\cite{bonacich2001eigenvector}, let $\{\mu_\ell\}$ and $\{\vec{u}_\ell\}$ be the (eventually multiple) eigenvalues and eigenvectors of the adjacency matrix $\bm{A}$, with $\widehat{\mu}= \mu_1 > \mu_\ell$ for $\ell\neq 1$. Then we can write $\bm{A} = \sum_{\ell=1}^N \mu_\ell \vec{u}_\ell \vec{u}_\ell^T$. Considering that $\bm{A}^k = \sum_{\ell=1}^N \mu_\ell^k \vec{u}_\ell \vec{u}_\ell^T $, from Eq.~\eqref{eq:alpha_centrality_def_2}  we have:
\begin{equation}\label{eq:alpha_centrality_eigenvector_centrality}
	\vec{c}^{\,(\alpha)} = \left(\sum_{k=0}^\infty \alpha^k \sum_{\ell=1}^N \mu_\ell^k \vec{u}_\ell \vec{u}_\ell^T \right)\vec{e} =
	\left(\sum_{\ell=1}^N \left(\sum_{k=0}^\infty \alpha^k  \mu_\ell^k\right) \vec{u}_\ell \vec{u}_\ell^T \right)\vec{e} = 
	\sum_{\ell=1}^N \frac{1}{1-\alpha\mu_\ell} \vec{u}_\ell \vec{u}_\ell^T \vec{e}.
\end{equation}
When $\alpha\to (1/\widehat{\mu})^-$, the factor relative to $\ell=1$ in the last term of Eq.~\eqref{eq:alpha_centrality_eigenvector_centrality} becomes the leading term, thanks to the Perron-Frobenius theorem, so that we can write:
\begin{equation}\label{eq:alpha_centrality_eigenvector_centrality_2}
	\lim_{\alpha\to(1/\widehat{\mu})^-}(1-\alpha_1)\vec{c}^{\,(\alpha)} =  (\vec{u}_1^T \vec{e}) \vec{u}_1 \propto \vec{u}_1 \propto \vec{c}^{\,(E)},
\end{equation}
where we have noted with $\vec{c}^{\,(E)}$ the eigenvector centrality.

Let us now generalize the analytical steps above to understand why the $\alpha$-centrality correlates with the fitted Heaps' exponents for generic graphs, as we showed numerically in the main text for real-world social networks. Let us suppose that the social network is a weakly-connected directed graph, since otherwise we can repeat the same argument for each weakly-connected component. As we have shown before, the asymptotic behavior of the Heaps' law for node $i$ is of the type $u_i \ln^{\widehat{p}(i)}(t) t^{\widehat{\lambda}(i)}$. We have shown also that the values of $\widehat{p}(i)$ and $\widehat{\lambda}(i)$ for each strongly connected component can be determined algorithmically. 
Here we will show that not only the $\alpha$-centrality can account for the coefficient $u_i$ like the eigenvector centrality, but also for the different values of $\widehat{p}(i)$ and $\widehat{\lambda}(i)$.
Let us first concentrate on what happens with multiple eigenvalues, for which the biggest difference is primarily given by $\widehat{p}(i)$. Therefore, let us suppose for now that all SCCs in the graph have the same Heaps' exponent $\widehat{\lambda}(i) = \widehat{\lambda}$, but different values of $\widehat{p}(i)$, and that in the leading terms the maximum value assumed by $\widehat{p}(i)$ is $\widehat{p}_{\max}<N$. This is the case for example of an open chain (already studied above), where $\widehat{p}(i)=0$, 1, \dots, $N-1$ for $i = N$, $N-1$, \dots, 1 respectively, and $\widehat{p}_{\max} = N-1$. Notice that, in this particular case, the adjacency matrix has only one eigenvalue $\widehat{\mu}$, related to the Heaps' exponent $\widehat{\lambda}$ through the relationship $\widehat{\lambda} = f(\widehat{\mu})$, with $f(x) = \frac{\nu}{\rho} + \frac{\nu+1}{\rho}x$. Therefore, the Jordan canonical form of the adjacency matrix is:
\begin{equation}\label{eq:adjacency_matrix_jordan_particular}
	\bm{A} = \bm{P} \bm{J} \bm{P}^{-1} = 
	\bm{P}\left[\begin{array}{ccccc}
		\widehat{\mu} & 1 & 0 & \cdots & 0\\
		0 & \widehat{\mu} & 1 & \ddots & \vdots \\
		\vdots & \ddots &  \ddots & \ddots & \vdots \\
		0 & \cdots & 0 & \widehat{\mu} & 1 \\
		0 & \cdots & 0 & 0 & \widehat{\mu} 
	\end{array}\right]\bm{P}^{-1} =
	\bm{P} (\widehat{\mu} \bm{J_0} + \bm{J_1}) \bm{P}^{-1},
\end{equation}
where $\bm{P} = \left[\begin{array}{c|c|c|c}
	\vec{u}_1 & \vec{u}_2 & \dots & \vec{u}_N
\end{array}\right]$ has the generalised eigenvectors in each column,
and $\bm{J_j}$ denotes the $N\times N$ matrix with ones only in the $(j+1)$-th upper diagonal and null everywhere else, with $\bm{J_0} = \bm{I}$.
%
It is possible to show that 
\begin{equation}
	\bm{A}^k = \bm{P}(\widehat{\mu}\bm{J_0} + \bm{J_1})^k \bm{P}^{-1} = \bm{P} \left( \sum_{j = 0}^{\min(N-1,\, k)} \binom{k}{j} \widehat{\mu}^{k-j}\bm{J_j} \right)\bm{P}^{-1} 
\end{equation}
Hence, from Eq~\eqref{eq:alpha_centrality_def_2}, similarly to what we have done in Eq.~\eqref{eq:alpha_centrality_eigenvector_centrality}, we have:
\begin{equation}\label{eq:alpha_centrality_general_calc}
	\begin{split}
		\vec{c}^{\,(\alpha)} &= \left(\sum_{k=0}^\infty \alpha^k \bm{P}(\widehat{\mu}\bm{J_0} + \bm{J_1})^k \bm{P}^{-1}\right)\vec{e} = 
		\left(\sum_{k=0}^\infty \alpha^k
		\bm{P} \left( \sum_{j = 0}^{\min(N-1,\, k)} \binom{k}{j} \widehat{\mu}^{k-j}\bm{J_j} \right)\bm{P}^{-1}  \right)\vec{e} =\\
		&=\left(\sum_{j = 0}^{N-1} \left(\sum_{k=j}^\infty \alpha^k
		\binom{k}{j} \widehat{\mu}^{k-j}\right) \bm{P}\bm{J_j}\bm{P}^{-1}  \right)\vec{e} 
		=\left(\sum_{j = 0}^{N-1} \left(\sum_{k=0}^\infty \alpha^{k+j}
		\binom{k+j}{j} \widehat{\mu}^{k}\right) \bm{P}\bm{J_j}\bm{P}^{-1}  \right)\vec{e} =\\
		&=\left(\sum_{j=0}^{N-1} \alpha^j\left(\sum_{k=0}^\infty \binom{k+j}{j} \alpha^k  \widehat{\mu} ^k\right) \sum_{\ell=1}^{N-j} \vec{u}_\ell\vec{u}_{\ell+j}^T \right)\vec{e} =
		\sum_{j=0}^{N-1} \sum_{\ell=1}^{N-j} \frac{\alpha^j}{(1-\alpha\widehat{\mu})^{j+1}} \vec{u}_\ell\vec{u}_{\ell+j}^T \vec{e} =\\
		&=  \sum_{\ell=1}^{N} \left(\sum_{j=0}^{N-\ell-1} \frac{\alpha^j}{(1-\alpha\widehat{\mu})^{j+1}} \vec{u}_{\ell+j}^T \vec{e}\right) \vec{u}_\ell.
	\end{split}
\end{equation}
From the above, it is clear that the nodes $\ell$ for which $(\vec{u}_1)_\ell$ is positive have the greatest $\alpha$-centrality when $\alpha\to1/\widehat{\mu}$, since they they are associated to the highest power in the logarithm $\widehat{p}(i) = \widehat{p}_{\max}$. Among these, as with the eigenvector centrality, nodes with higher coefficients (corresponding to the eigenvector centralities in that SCC) have higher ranking. Then the nodes who have zeroes in $\vec{u}_1$ but positive entries in $\vec{u}_2$ are next in the ranking, and so on. This confirms the fact that, when comparing nodes with same asymptotic Heaps' exponent, those with higher discovery rates, i.e. those with higher powers in the logarithm factor, have the highest $\alpha$-centrality. 

\medskip

A similar approach to the one we used to derive the algorithmic solution of the Heaps’law for a generic graph can be used to treat generic weakly-connected graphs.
Let us divide the network into its SCCs. For each component $C_q$, we denote $\mu^{(q)}$ the maximum between the maximum eigenvalue the component would have if isolated and the maximum eigenvalue of the neighboring SCCs, following the same order used with the developed algorithm. In this setting, it is then possible to compute the $\alpha$-centrality at $\alpha\to(1/\mu^{(q)})^-$, that might be different across SCCs. The final ranking is given by ordering the evaluated $\alpha$-centralities starting from those with the highest $\mu^{(q)}$.

\begin{figure}[htbp]
	\includegraphics[width=0.8\textwidth]{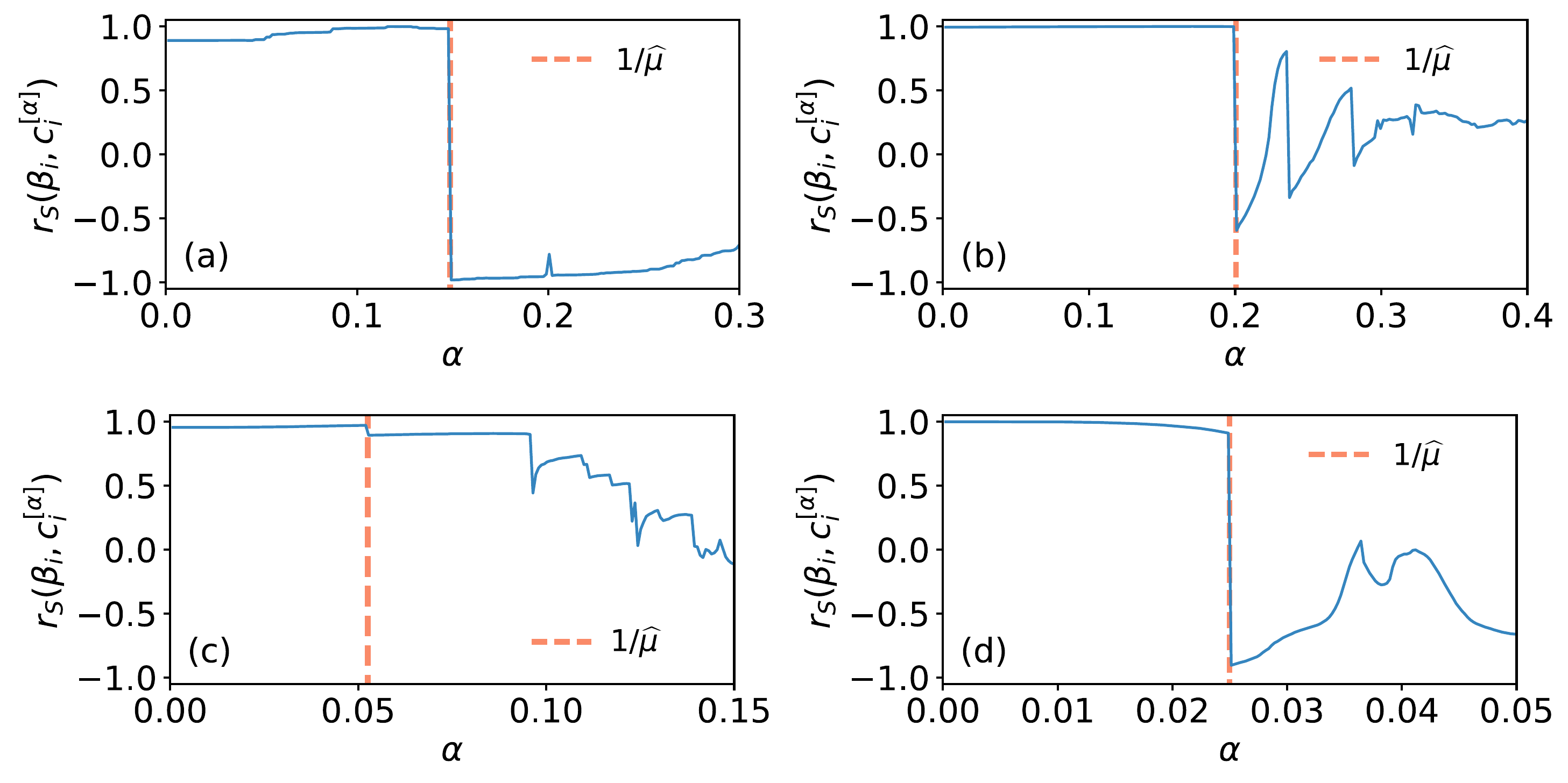}
	\caption{Spearman's rank correlation $r_{S}$ between paces of discovery $\beta_i(10^4)$ and $\alpha$-centrality $c^{[\alpha]}_i$ as a function of $\alpha$ for nodes $i=1,\dots,N$ belonging to four different real-world networks: (a) the Zachary Karate Club network \cite{zachary1977information}, (b) a network of follower relationships of Twitter \cite{de2010does}, (c) a co-authorship network in network science \cite{newman2006finding} and (d) a collaboration network between jazz musicians \cite{gleiser2003community}. Each dashed vertical line corresponds the value of $1/\widehat{\mu}$, with $\widehat{\mu}$ denoting the maximum eigenvalue of the corresponding adjacency matrix. The parameters of the model are $\rho=10$, $\nu=1$, $M_0=\nu+1$.
	}
	\label{fig:SI_spearman_corr_vs_alpha}
\end{figure}

It is worth noticing that this method can be computationally not efficient, especially for big networks. For this reason, we have tested how reliable the $\alpha$-centrality with the same value of $\alpha$ is when comparing it to the Heaps' exponents, regardless of the procedure above.
In Fig.~4 of the main text we have investigated the relation between Heaps' exponents and $\alpha$-centralities setting $\alpha$ to $0.85/\widehat{\mu}$. Here, we further investigate how the correlation changes with $\alpha$. This is shown in Fig.~\ref{fig:SI_spearman_corr_vs_alpha}, where we plot the Spearman's rank correlation coefficient between the paces of discovery $\beta_i(10^4)$ and the $\alpha$-centralities $c_i^{[\alpha]}$ as a function of $\alpha$ for all the nodes $i=1,\dots,N$ composing the four considered real-world networks. Although panel (d) displays a decrease in the correlation when approaching $1/\widehat{\mu}$, however, setting $\alpha < 1/\widehat{\mu}$ leads to Spearman's rank correlation coefficients $r_S>0.89$ in all four cases (cfr main text). 

\end{document}